\documentclass[10pt,twocolumn]{IEEEtran}
\usepackage{xspace,amsthm,amsmath,amssymb,amsfonts,epsfig,syntonly}
\usepackage{cite,bm,color,url,textcomp,balance}
\usepackage{epstopdf}
\usepackage{empheq}

\usepackage{algorithm}
\usepackage{algorithmicx}
\usepackage{algpseudocode}

\usepackage{graphicx}{\large}
\usepackage{subfigure}
\usepackage{booktabs, multicol, multirow}

\usepackage{amsmath,amssymb,amsthm, enumerate}
\usepackage{bigints}
\usepackage{stfloats,siunitx}

\newcommand*{\red}{\color{black}}
\newcommand*{\blue}{\color{black}}

\DeclareMathOperator*{\argmin}{arg\,min}

\long\def\symbolfootnote[#1]#2{\begingroup
\def\thefootnote{\fnsymbol{footnote}}
\footnote[#1]{#2}\endgroup}

\psfull

\allowdisplaybreaks[4]

\IEEEoverridecommandlockouts

\begin{document}
\title{Joint Optimization of Trajectory, Propulsion and Thrust Powers for Covert UAV-on-UAV Video Tracking and Surveillance}

\author{Shuyan Hu,~\IEEEmembership{Member, IEEE}, Wei Ni,~\IEEEmembership{Senior Member, IEEE}, Xin Wang,~\IEEEmembership{Senior Member, IEEE}, \\ Abbas Jamalipour,~\IEEEmembership{Fellow, IEEE},
and Dean Ta,~\IEEEmembership{Member, IEEE}
\thanks{S. Hu, X. Wang and D. Ta are with the School of Information Science and Technology, Fudan University, Shanghai 200433, China (e-mails: syhu14@fudan.edu.cn; xwang11@fudan.edu.cn; tda@fudan.edu.cn).

W. Ni is with the Data61, Commonwealth Scientific and Industrial Research Organization, Sydney, NSW 2122, Australia (e-mail: wei.ni@data61.csiro.au).

A. Jamalipour is with the School of Electrical and Information Engineering, The University of Sydney, Sydney, NSW 2006, Australia
(email: a.jamalipour@ieee.org)

(c) 2020 IEEE. Personal use is permitted, but republication/redistribution requires IEEE permission.

}}

\markboth{ACCEPTED BY IEEE TRANSACTIONS ON INFORMATION FORENSICS AND SECURITY}%
{HU \MakeLowercase{\textit{et al.}}: Joint Optimization of Trajectory, Propulsion and Thrust Powers for Covert UAV-on-UAV Video Tracking and Surveillance}

\maketitle

\begin{abstract}
Autonomous tracking of suspicious unmanned aerial vehicles (UAVs) by legitimate monitoring UAVs (or monitors)
can be crucial to public safety and security. 
It is non-trivial to optimize the trajectory of a monitor while conceiving its monitoring intention,
due to typically non-convex propulsion and thrust power functions.
This paper presents a novel framework to jointly optimize the propulsion and thrust powers, as well as the 3D trajectory of a solar-powered monitor which conducts covert, video-based, UAV-on-UAV tracking and surveillance.
A multi-objective problem is formulated to minimize the energy consumption of the monitor
and maximize a weighted sum of distance keeping and altitude changing,
which measures the disguising of the monitor.
Based on the practical power models of the UAV propulsion, thrust and hovering, and the model of the harvested solar power,
the problem is non-convex and intangible for existing solvers.
We convexify the propulsion power by variable substitution, and linearize the solar power.
With successive convex approximation, the resultant problem is then transformed with tightened constraints and efficiently solved by the proximal difference-of-convex algorithm with extrapolation in polynomial time.
The proposed scheme can be also applied online.
Extensive simulations corroborate the merits of the scheme, as compared to baseline schemes with partial
or no disguising.

\end{abstract}


\section{Introduction}

With their excellent flexibility, swift deployment, and wide coverage,
unmanned aerial vehicles (UAVs) have been extensively applied to search and rescue \cite{alota19},
landscape or power line inspection \cite{zhou18},
bushfire monitoring \cite{chi19}, parcel delivery~\cite{wang19parcel},
data dissemination \cite{zeng19}, relaying~\cite{kli16},
eavesdropping~\cite{tang19, yuan20, hu20},
and mobile edge computing \cite{yang19}.
Equipped with sensors and cameras, UAVs have also been deployed to track and monitor mobile targets \cite{hailong20tii}.
The UAVs can record the misbehaviors of the targets during tracking as evidence for forensics purpose~\cite{fanid20}.
However, the targets may notice the presence of the monitor, and carry out countermeasures,
change their behaviors and potentially endanger public safety.
In this sense, the covertness of the UAVs' operation is crucial, especially for public safety reasons.

Given a resolution of on-board cameras, a UAV carrying out covert video surveillance has to keep the target within its visual range,
while maintaining a distance from the target for covertness.
Visual-based control methods are developed to enable a UAV to track a moving ground target
\cite{zhang20, hailong20, hailong19cn, huang16},
estimate the three-dimensional (3D) geolocation of a moving target~\cite{zhang18tie},
or search for a stationary target \cite{yao19}.
In \cite{zhang20}, the nonlinear hierarchical control was integrated with the geometric control, where a
visual servoing strategy was proposed to remove the need of a measurable thrust force or its derivative,
compared with typical backstepping methods, greatly simplifying the practical implementations.
Based on Lyapunov techniques, the vision-based controller was proved to be asymptotically stable.
Target tracking by UAVs in urban environments was studied in \cite{shafer08} and \cite{yu15} by considering vision occlusion.
The UAV trajectory was designed to maximize the probability of detection, i.e., keeping the target inside the field of view (FoV).
A dynamic zero-sum game was proposed in \cite{quintero14} to optimize the heading of the monitor over a finite period,
where the monitor wishes to keep the target within its proximity and visibility.
Dynamic programming was used to minimize the expected cumulative cost
depending on the monitor's distance from the target.
{\blue
Among the works on UAV-based target tracking, e.g., \cite{zhang20}--[21],
3D UAV trajectory design was only considered in \cite{hailong19cn}, and the rest assumed that the UAV flies at a constant altitude
and only planned 2D (horizontal) UAV trajectories.
In \cite{hailong19cn}, the energy consumption of a surveillance UAV was minimized,
while the number of observable targets was maximized.}
Moreover, none of these works have considered the covertness or stealth of the monitor during the tracking or surveillance processes.

A separate challenge of covert video tracking and surveillance is the finite capacity of the on-board battery.
The trajectory of the UAV has to be meticulously planned to extend the duration of the target monitoring,
as suggested in~\cite{hailong19cn, hailong20, huang16}.
Practical flight power consumption models capturing different flight modes, e.g., propulsion, thrust and hovering, need to be taken into account.
Yet, the power consumption of the UAV was simply modeled as a linear function of its speed in \cite{hailong19cn, hailong20},
while the power consumption of the hovering mode was not considered in \cite{huang16}.

Solar power has been increasingly utilized in UAV platforms, e.g., \cite{brizon}.
Many latest UAVs can carry a payload of as much as $10$ kg.
It is plausible for such UAVs to be equipped with solar panels.
The harvested solar power can potentially energize the UAVs' tracking and surveillance missions for sustainability and longevity.
However, it is non-trivial to plan the trajectories of solar powered UAVs
due to the non-convexity of the power harvesting process~\cite{aglie}.

{\red
In a different context, 3D trajectory plans have been studied for UAV communications with the objective of throughput maximization,
where the UAV trajectory, transmit power, and resource allocation were jointly optimized~\cite{derrick, hua, wmfeng}.
In~\cite{derrick}, a solar-powered UAV was employed to provide wireless communication services to multiple ground users.
Multiple UAVs were deployed for simultaneous data dissemination and collection from sensor nodes~\cite{hua},
or emergency communication to internet-of-things devices in a disaster area~\cite{wmfeng}.}
In the presence of eavesdroppers, jamming-aided secure UAV-assisted communications were pursued in \cite{cai} and \cite{ruide}.
Given the altitude of a UAV transmitter, user scheduling, transmit power, jamming policy, and the 2D trajectory of the UAV
were jointly optimized to maximize the system energy efficiency while guaranteeing the quality-of-service of the users in \cite{cai},
and to maximize the average minimum secrecy rate per user in \cite{ruide}.
{\red Intelligent reflecting surface (IRS)-aided secure UAV communication was investigated in \cite{ssfang},
where the secrecy rate of a UAV transmitter was maximized by jointly optimizing the power control and 2D trajectory of the UAV,
and the phase shifts of the IRS.}
However, all the 3D trajectory planning techniques are not directly applicable to covert UAV-on-UAV video surveillance,
due to distinctive system settings.

This paper presents a new approach to covert (or disguised) UAV-on-UAV, visual surveillance,
where a solar-powered, rotary-wing, monitoring UAV (or ``monitor'')
equipped with video cameras tracks, follows, and visually monitors a suspicious UAV (or ``target'').
The monitor disguises its intention of monitoring by adjusting its trajectory and heading,
so that it appears to be far away and fly randomly in the view of the target
and does not draw the target's attention.
The monitor also adapts its trajectory and heading to the solar power harvesting process,
thereby extending the mission time and sustainability.
Such UAV-based video surveillance can have important applications to public safety and security.

The considered scenario is new and, to the best of our knowledge, has yet to be rigorously investigated in the existing literature
despite its practical importance.
In particular, the target is mobile in the new scenario, while it is often stationary in the existing studies \cite{yao19}.
Moreover, visual disguise is considered in the new scenario, which has not been well investigated in the existing studies.
All of these require new modeling and solutions for the new scenario.
The key contributions of this paper are summarized as follows.
\begin{itemize}
\item[$\bullet$] To optimize the trajectory and heading of the monitor, a new problem is formulated to account for
both monitoring and disguising in the solar-powered, covert, UAV-on-UAV, video surveillance.
The disguising is measured by the monitor's distance to the target and its altitude changes.
The trajectory and heading are also adjusted adapting to the solar power harvesting process.
\item[$\bullet$] Practical, non-convex UAV propulsion power and solar power models are captured in the constraints,
and convexified by iteratively linearly tightening. The resulting problem becomes a difference-of-convex (DC) program.
Provided the target's trajectory, a suboptimal solution is efficiently obtained
with polynomial time-complexity by leveraging the proximal difference-of-convex
algorithm with extrapolation (PDCAE) method.
\item[$\bullet$] The proposed approach is generalized online, where the approach serves as
the dynamic control method of the monitor and specifies the trajectory (one waypoint at a time) on-the-fly,
e.g., following the model-predictive control.
The disguising is further enhanced online by misaligning the headings of the monitor and target.
\end{itemize}
\noindent Extensive simulations corroborate the merits of our scheme over the baseline schemes with partial
(i.e., only distance keeping or altitude changing) or no disguising.

The rest of the paper is organized as follows.
Section \ref{sec.model} describes the system models.
Section \ref{sec.formulate} formulates the problem of covert video tracking and surveillance by a solar-powered UAV.
Section \ref{sec.solution} delineates our approach to solve the problem of joint power and 3D trajectory optimization
offline, when the movement of the target is known in prior.
The extension of the framework to the online application is described in Section \ref{sec.extend}.
Numerical results are provided in Section \ref{sec.sim},
followed by a conclusion in Section~\ref{sec.con}.

\emph{Notation}: $\mathbb R^n$ denotes the $n$-dimensional Euclidean space;
$\langle \cdot , \cdot \rangle$ and $|| \cdot ||$ stand for the inner product and Euclidean norm, respectively;
dom\,$F$ is the domain of function $F$;
$\partial F$ denotes the sub-differential of function $F$;%
\footnote{For a convex function $F$ defined on ${\mathbb R}^n$, its sub-differential at $x_0 \in {\mathbb R}^n$ is defined as
$\partial F(x_0):=\{v \in {\mathbb R}^n | F(x)-F(x_0) \ge \langle v, x-x_0 \rangle, \forall x \in {\mathbb R}^n \}$.}
and $\nabla F$ stands for the gradient of a continuously differentiable function $F$.
The notations used in the paper are listed in Table \ref{tab.list}.

\begin{table}[t]
\caption{A list of notations and variables.}
\begin{center}
\begin{tabular}{l l}
\toprule
Notation                 & Description \\ \midrule

$\cal F$                & A set defining the feasible flight region of the monitor \\
$T, ~N$             & Scheduling period and total number of time slots \\
$\delta$                 &Duration of each time slot \\
$(x_0,y_0,z_0)$    &Initial waypoint of the monitor \\
$V_{hm},~V_{vm}$       & Maximum horizontal and vertical speeds of the monitor \\
$P_0$, $P_1$              &Blade profile power and induced power \\ 
$s$, $A$                      &Rotor solidity and disc area \\ 
$U_{tip}$     &Tip speed of the rotor blade \\ 
$v_0$      &Mean rotor induce velocity \\ 
$\rho$, $d_f$      &Atmospheric density and fuselage drag ratio \\
$W$     & Body weight of the monitor \\
$E_0$   & Amount of initial energy in the battery \\
$\eta_0$ & Ratio of usable energy in the battery \\
$\eta$, $S$          &Efficiency and size of a solar panel \\
$P_i$          & Power intensity of solar beams \\
$\alpha$           & Sum atmospheric extinction \\
$z_l$                       & Minimum altitude of the monitor \\             
$D$                       & Maximum 3D monitor-target distance   \\
$c_1,~c_2$          & Coefficients for solar power approximation \\
$\mu_1,~\mu_2$   & Weights for disguising \\
$t$, $\ell$                      &Index of each time slot and each iteration \\
$(a_t, b_t,H)$    & 3D waypoints of the target at time slot $t$ \\
$(x_t,y_t,z_t)$   & 3D waypoints of the monitor at time slot $t$ \\
\multirow{2}*{$\Delta_t$}        & Altitude change of the monitor between time slots $t$ \\ 
~&  and $t-1$\\
$P_h^t,~P_v^t$   & Propulsion and thrust power of the monitor at time slot $t$ \\ 
$P_s^t,~\hat P_s^t$  & Harvested solar power and its approximation at time slot $t$ \\
$\vartheta_t$ & Solar zenith angle at time slot $t$ \\
$d_t,~\hat d_t$ & 2D and 3D monitor-target distances at time slot $t$ \\
$f_t$                 &Performance of disguising at time slot $t$ \\
$q_t$                & Auxiliary variable at time slot $t$ \\
$\boldsymbol \lambda_{(\ell)}$  &Extrapolation of all the variables at iteration $\ell$ \\
$\beta_{\ell}$    & Extrapolation parameter at iteration $\ell$ \\
\bottomrule
\end{tabular}
\end{center}
\label{tab.list}
\end{table}

\section{System Model}\label{sec.model}

\begin{figure}[t]
\centering
\includegraphics[width=0.47\textwidth]{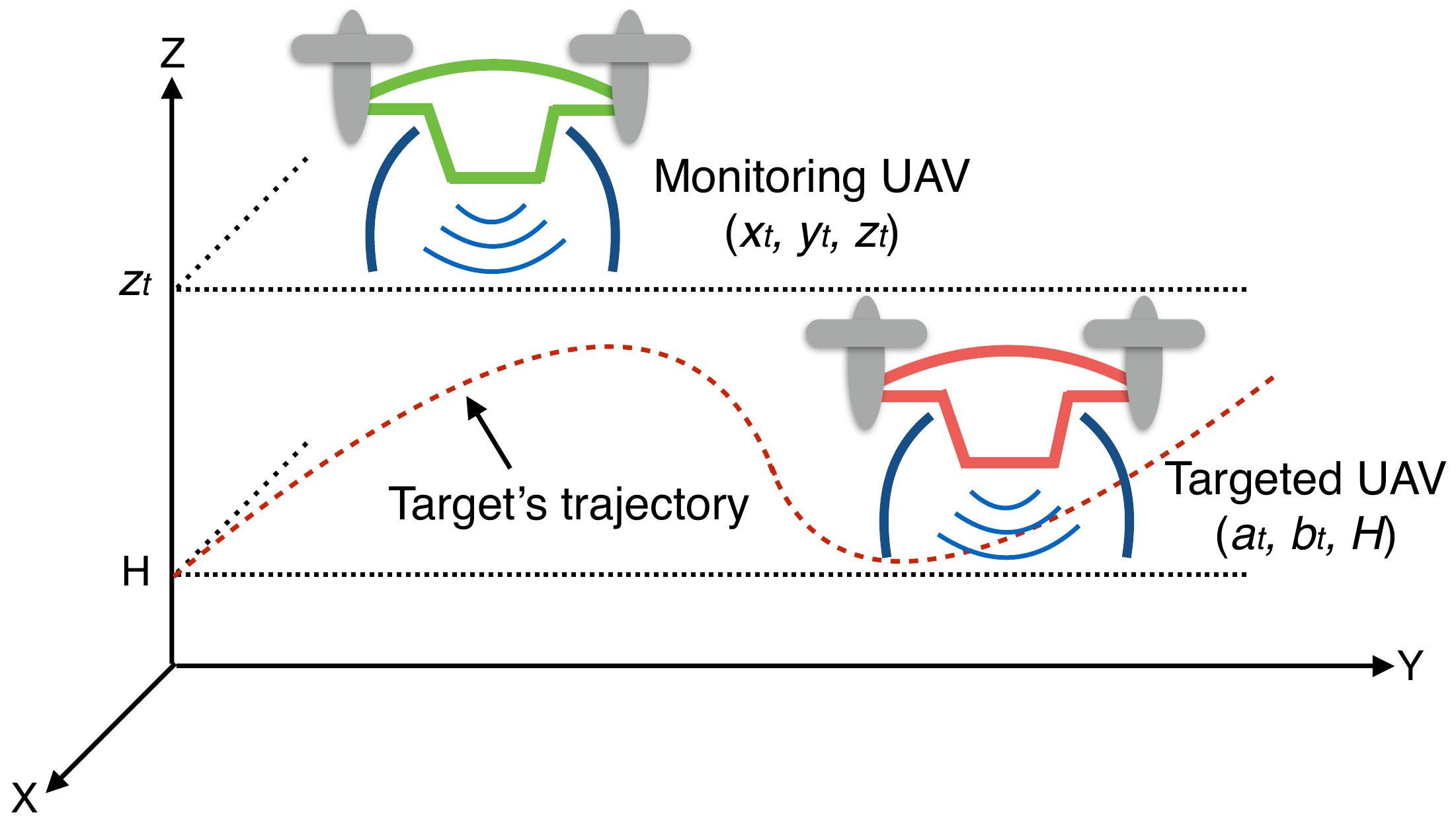}
\caption{A UAV-enabled video surveillance system.}
\label{systmod}
\end{figure}

In the considered system, a rotary-wing monitoring UAV with a 360-degree panoramic camera is employed to track and monitor a suspicious target UAV traveling at a fixed altitude of $H$ (in meters).
We assume that the movement of the target is perfectly predictable;
see Fig. \ref{systmod}.
The UAV can move forward horizontally and vertically, or hover.
The monitoring UAV does not have a specific destination.
It is dispatched on-spot once the target is detected,
and does not fly back to its home base until the surveillance mission is completed.

\subsection{UAV Mobility}\label{sec.uavmob}
We consider a 3D Cartesian coordinate system.
The monitoring mission lasts a finite period of $T$ seconds.
We divide the period $T$ evenly into $N$ time slots , i.e., $t=1,\cdots,N$.
The duration of a slot, denoted by $\delta$,
is so short that the position of the UAVs during each slot can be indicated as a single waypoint.
The waypoints of the monitor, denoted by $(x_t,y_t,z_t), \forall t=1,\cdots,N$, are 3D.
We suppose that the initial location of the monitor is $(x_0, y_0, z_0)= (0, 0, z_0)$, when the video surveillance starts,
with $z_0 > H$.
The maximum horizontal and vertical speeds of the monitor are $V_{hm}$ and $V_{vm}$, respectively.
The altitude change is $\Delta_t$ between two successive time slots $t-1$ and $t$.
The monitor's mobility constraints, including its initial location and speed constraints, are given by~\cite{zeng16, derrick}:
\begin{subequations}
\begin{align}
(x_1-0)^2+(y_1-0)^2 &\le (V_{hm} \delta)^2, \label{eq.mob1} \\
\Delta_1 := | z_1-z_0 | &\le V_{vm} \delta, \label{eq.mob2} \\
(x_t-x_{t-1})^2+(y_t-y_{t-1})^2  &\leq (V_{hm} \delta)^2, ~\forall t,  \label{eq.mob3} \\
\Delta_t := | z_t-z_{t-1} | &\leq V_{vm} \delta, ~\forall t,  \label{eq.mob4}
\end{align}
\end{subequations}
where $V_{hm} \delta$ and $V_{vm} \delta$ are the largest horizontal and vertical distances
that the monitor can travel between two consecutive slots (i.e., two consecutive waypoints), respectively.

\subsection{UAV Propulsion and Thrust Powers}

For a rotary-wing UAV with a horizontal speed $V_h^t$,
the propulsion power at time slot $t$, denoted by $P_h^t$, is given by~\cite{zyong}
\begin{equation}\label{speed}
\begin{aligned}
P_h^t = & P_0 \left(1+\frac{3{V_h^t}^2}{U_{tip}^2}\right) + P_1\left(\sqrt{1+\frac{{V_h^t}^4}{4v_0^4}}
-\frac{{V_h^t}^2}{2 v_0^2}\right)^{\frac{1}{2}}\\
&+ \frac{1}{2} d_f\rho s A{V_h^t}^3,
\end{aligned}
\end{equation}
where $P_0$ and $P_1$ are the constant {\it blade profile power} and {\it induced power} in a  hovering mode,
respectively; $U_{tip}$ is the tip speed of the rotor blade; $v_0$ denotes the average rotor-induced velocity in the hovering mode;
$d_f$ and $s$ denote the fuselage drag ratio and rotor solidity, respectively;
and $\rho$ and $A$ are the atmospheric density and rotor disc area, respectively.

When $V_h^t=0$, Eq. \eqref{speed} is the power consumption of the hovering state.
We suppose that within each slot $t$, the UAV maintains a constant speed, as given by
\begin{equation}\label{vt}
V_h^t = \sqrt{(x_t-x_{t-1})^2 + (y_t - y_{t-1})^2}/ \delta, ~\forall t.
\end{equation}
By substituting \eqref{vt} into \eqref{speed}, we find that the first and the third terms of \eqref{speed}
are jointly convex with respect to $(x_t,x_{t-1}, y_t, y_{t-1})$.
However, the second term is neither convex nor concave.

The thrust power of vertical flight, i.e., altitude change, can be obtained by \cite[eq. (6)]{derrick}, \cite[eq. (7.12)]{seddon11}
\begin{equation}\label{eq.thrust}
P_v^t = W(z_t - z_{t-1})/ \delta, ~\forall t,
\end{equation}
where $W$ is the body weight of the UAV.
Clearly, $P_v^t$ is jointly convex with respect to $(z_t,z_{t-1})$.
The thrust power is positive for (altitude) ascending,
and it is negative for (altitude) descending since energy is saved thanks to gravity~\cite{derrick}.

\subsection{Harvested Solar Power}
Suppose that the monitor has a solar panel to harvest energy and a rechargeable battery to store energy.
The battery is initially charged energy of $E_0$ (in Joules).
The monitor needs to maintain a required minimum energy reserve $(1-\eta_0)E_0$ at any moment for
sustaining operations and dealing with emergency.
In practice, the monitor flies within a certain altitude range for aviation control and collision avoidance.
In the scenario of tracking and monitoring for civilian purpose,
the monitor typically flies lower than $1000$ m.
In this range of altitude, the harvested solar power of the monitor at each time slot $t$ is given by~\cite{aglie}
\begin{equation}\label{eq.solar}
P_s^t(z_t)= \eta SP_i \exp\left(-\frac{\alpha}{ \cos \vartheta_t}\left(1-2.2556\times 10^{-5} z_t  \right)^{5.2561} \right),
\end{equation}
where $\eta \in (0,1)$ and $S$ (in m$^2$) denote the efficiency and size of the solar panel, respectively;
$P_i$ (in Watts) is the constant power intensity of the solar beams before entering the atmosphere;
$\alpha>0$ is the sum atmospheric extinction;
and $\vartheta_t \in [0,  \pi/2]$ denotes the solar zenith angle at slot $t$.
Note that $\vartheta_t$ is $\pi/2$ at sunrise (or sunset), and then gradually decreases to $0$ at midday.
The solar zenith angle changes negligibly within half an hour.
Since the surveillance mission lasts much shorter than half an hour in the investigated case,
$\vartheta_t$ is assumed to remain unchanged during the period of $T$.

Without loss of generality (w.l.o.g.), we assume that the surveillance mission takes place at the midday of a sunny day
and therefore $\cos(\vartheta_t) =1$.
At each time slot $t$, the harvested solar power \eqref{eq.solar} is simplified as
\begin{equation}\label{eq.solar2}
P_s^t(z_t)= \eta SP_i \exp\left(-\alpha \left(1-2.2556\times 10^{-5} z_t  \right)^{5.2561} \right).
\end{equation}
Nevertheless, the approach proposed in this paper can be applied to tackle time-varying solar zenith angle
for a much longer period.

\subsection{Visual Disguise}\label{sec.disguise}

Let $(a_t, b_t, H)$ denote the 3D waypoints of the target at time slot $t$.
The time-varying distances between the monitor and the target on the horizontal $(x,y)$-plane
and in the 3D space are given by
\begin{equation}\label{eq.dist2}
d_t = \sqrt{(x_t-a_t)^2 + (y_t - b_t)^2}, ~\forall t;
\end{equation}
\begin{equation}\label{eq.distance}
\hat d_t = \sqrt{(x_t-a_t)^2 + (y_t - b_t)^2 + (z_t-H)^2}, ~\forall t.
\end{equation}

We assume that the monitor always flies horizontally behind and above the target,
i.e., $x_t \le a_t$, $y_t \le b_t, \forall t$,%
\footnote{Our proposed scheme can be readily applied to the problem without the requirements of $x_t \le a_t$ and $y_t \le b_t, \forall t$.}
and $z_t \ge z_l >H, \forall t$,
where $z_l$ provides the lower-bound altitude of the monitor.
This is because the panoramic camera is typically installed at the bottom of the monitor.
In order to successfully track the target, the monitor has to travel inside a {\it feasible flight region} (FFR)
specified by the above constraints:
\begin{align}\label{eq.ffr}
{\cal F} := \{(x_t, y_t, z_t) | & (\hat d_t)^2 \le D^2, x_t \le a_t, y_t \le b_t, \notag \\
&  H < z_l \le z_t, ~\forall t\},
\end{align}
where $D$ specifies the distance for effective visual surveillance given the resolution of the panoramic camera at the monitor.
The upper limit of the monitor's altitude is bounded by \eqref{eq.distance}.

{\blue
Different from the hypothesis test extensively considered in covert wireless communication,
the monitor delivers covertness by visual disguise.
The reason for the consideration of visual disguise is because optical cameras (e.g., including those at the target)
can often achieve persistent observations if unobstructed,
given the infinite maximum focusing distances of typical camera lens \cite{liucamera}.

We propose that the monitor disguises by keeping its distance from the target (at a higher altitude),
e.g., \eqref{eq.dist2}, and constantly changing its altitude $\Delta_t$ to confuse the target, as considered in \cite{hailong20}.
The constantly changing altitude of the monitor makes it difficult for the target to correctly focus its camera lens on the monitor for a clear view of the monitor.
Keeping the monitor's distance from the target makes it hard for the target to get good image resolutions on the monitor, even if its camera is correctly focused.
}

\section{Problem Formulation}\label{sec.formulate}

In this section, we pursue the optimal trajectory and power management of the solar-powered rotary-wing monitor,
first offline by minimizing the total energy consumption and maximizing the disguising performance
of the monitor.
The offline scheme can be implemented if:
i) the target's route is perfectly predictable, e.g., based on historical data;
or ii) the target has only one possible trajectory to travel for the considered period
(which is only part of the surveillance mission).
The offline scheme can be extended for online operations, as will be discussed in Section \ref{sec.extend}.

The monitor can fly horizontally and vertically to adjust its position or posture
for tracking, monitoring, and disguising.
We use $\mu_1 d_t^2$ and $\mu_2 \Delta_t^2$ to measure the disguising performances of the monitor
in regards to (horizontal) distance keeping and altitude changing at each time slot, respectively.
Here, $\mu_1$ and $\mu_2$ (in Watts/m$^2$) are two nonnegative coefficients weighting the two aspects.
The reason for decoupling the disguising measure between the horizontal and vertical controls is due to the fact
that the propulsion and thrust of a UAV are typically controlled separately.
The propulsion and thrust powers yield different models specified in \eqref{speed} and \eqref{eq.thrust}, respectively.
Note that $\mu_1 d_t^2$ is jointly convex in $(x_t, y_t)$,
and $\mu_2 \Delta_t^2$ is continuously differentiable and jointly convex in $(z_t, z_{t-1}), \forall t$.

Given the monitor's and the target's waypoints $(x_t, y_t, z_t)$ and $(a_t, b_t, H)$ at the $t$-th time slot,
the propulsion and thrust powers of the monitor $P_h^t$ and $P_v^t$,
its harvested solar power $P_s^t$,
disguising measure $f_t := \mu_1 d_t^2 + \mu_2 \Delta_t^2$,
and mobility constraints \eqref{eq.mob1}--\eqref{eq.mob4},
the problem of interest is cast as
\begin{subequations}\label{p1}
\begin{align}
&\min_{\{x_t, y_t, z_t, \forall t \}} \sum_{t =1}^N (P_h^t+P_v^t - f_t) \delta \label{p11}\\
\text {s.t.} ~&\sum_{n=1}^t (P_h^n + P_v^n ) \delta \leq \sum_{n=1}^t  P_s^n(z_n) \delta + \eta_0 E_0, ~\forall t, \label{p12}\\
&| z_t- z_{t-1} - (z_{t-1} - z_{t-2}) | \le V_{vm} \delta, \forall t \ge 2, \label{p14} \\
& (x_t, y_t, z_t) \in {\cal F}, ~\forall t, \label{p13}\\
&\eqref{eq.mob1} - \eqref{eq.mob4}, \notag
\end{align}
\end{subequations}
where \eqref{p12} is an energy harvesting causality constraint
to ensure that the total energy consumed by a time slot
does not exceed the total of the energies initially charged and progressively harvested by the time slot,
while the monitor can still maintain a minimum energy reserve of $(1-\eta_0)E_0$ Joules at any time slot for other functionalities;
\eqref{p14} indicates that the altitude change between two consecutive slots does not exceed the maximum amount;
and $f_t$ is jointly convex with respect to $(x_t, y_t, z_t, z_{t-1})$, $t=1,\cdots,N$.

{\blue
We note that the properties of video tracking are parameterized in \eqref{p11} and \eqref{p13}.
Specifically, \eqref{eq.distance} and \eqref{eq.ffr}, and in turn, \eqref{p13},
indicate that the monitor needs to track and follow the target by keeping the target within its sight.
The maximum visual range of the monitor can be specified in prior,
depending on the optical characteristics of the monitor's camera lens
(e.g., focal length, zoom capability, zoom magnification, etc.), the illumination condition of the environment~\cite{schramm},
and the resolution requirement on the target.

The parameter $f_t$ in \eqref{p11} specifies that the monitor needs to keep its distance from the target
(as long as the target is within its sight), and also keep changing its altitude.
As mentioned in Section \ref{sec.disguise}, the constantly changing altitude of the monitor makes it difficult for the target to correctly focus its camera lens on the monitor for a clear view of the monitor.
Keeping the monitor's distance from the target makes it hard for the target to get good image resolutions on the monitor,
even if correctly focused.
By this means, the properties of video tracking are captured in problem \eqref{p1}.
}

Problem \eqref{p1} is not convex, due to the non-convex term in $P_h^t$,
the minimization of a concave function $-f_t$,
and the non-convex element $P_s^t$ in constraint \eqref{p12}.
Therefore, it is difficult to tackle \eqref{p1} with standard convex solvers,
such as the interior point method~\cite{Boyd}.
In the next section, we propose a new efficient method to convexify the non-convex terms in \eqref{p1}
and solve the problem with fast polynomial-time convergence.

\section{Proposed Algorithm for Solar-powered Covert UAV-on-UAV Video Surveillance}\label{sec.solution}
In this section, we leverage the SCA \cite{sca18} and PDCAE~\cite{wen18}
to tackle the non-convexity of \eqref{p1}, and obtain a low-complexity solution for \eqref{p1}.
The SCA technique successively approaches the global upper bound of the propulsion power $P_h^t$.
The PDCAE replaces the concave part of the objective \eqref{p11}, i.e., $-f_t$,
with a linear function and solves the resultant convex problem.
Non-trivial mathematic manipulations of \eqref{p1} are involved to orchestrate SCA and PDCAE.

\subsection{SCA-based Convexification}

We start with the non-convex part of $P_h^t$, i.e.,
$\left(\sqrt{1+\frac{{V_h^t}^4}{4v_0^4}}-\frac{{V_h^t}^2}{2 v_0^2}\right)^{\frac{1}{2}}$ in \eqref{speed},
by defining new slack variables $\{q_t \geq 0,\, \forall t =1, \cdots, N\}$:
\begin{equation}
q_t^2 = \sqrt{1+\frac{{V_h^t}^4}{4v_0^4}}-\frac{{V_h^t}^2}{2 v_0^2}, ~\forall t,
\end{equation}
which can be reorganized as
\begin{equation}\label{mu}
\frac{1}{q_t^2} = q_t^2 + \frac{{V_h^t}^2}{v_0^2}, ~\forall t.
\end{equation}
As a result, the second term on the right-hand side (RHS) of \eqref{speed} can be substituted by the linear component $P_1 q_t$,
with an additional new constraint \eqref{mu}.

For the purpose of exposition, we now integrate the expression for $V_h^t$ in \eqref{vt} and define
\begin{equation}\label{newpm}
\begin{aligned}
{\tilde P}_h^t := & P_0 + \frac{3P_0}{U_{tip}^2 \delta^2}\left[(x_t - x_{t-1})^2 + (y_t - y_{t-1})^2\right] + P_1 q_t \\
& + \frac{d_f}{2\delta^3} \rho s A\left[(x_t - x_{t-1})^2 + (y_t - y_{t-1})^2\right]^{3/2}, ~\forall t.
\end{aligned}
\end{equation}
Here, ${\tilde P}_h^t$ is jointly convex in $(x_t,x_{t-1}, y_t, y_{t-1}, q_t)$.
Given $\delta$, problem \eqref{p1} can be rewritten as
\begin{subequations}\label{p2}
\begin{align}
&\min_{\{x_t, y_t, z_t, q_t, \forall t \}} \sum_{t=1}^N (\tilde P_h^t+P_v^t - f_t) \label{p21}\\
\text {s.t.} ~&\sum_{n=1}^t (\tilde P_h^n + P_v^n ) \leq \sum_{n=1}^t P_s^n(z_n)  + \eta_0 E_0/ \delta, ~\forall t, \label{p22} \\
&\frac{1}{q_t^2} \leq q_t^2 + \frac{(x_t - x_{t-1})^2 + (y_t - y_{t-1})^2}{\hat v_0^2 }, ~\forall t, \label{p23}\\
&\eqref{eq.mob1} - \eqref{eq.mob4}, \eqref{p14}~\text{and}~\eqref{p13}, \notag
\end{align}
\end{subequations}
where $\hat v_0^2=v_0^2 \delta^2$.
Constraint \eqref{p23} is obtained by relaxing the equality in \eqref{mu} with inequality.
Yet, the equivalence still holds between problems \eqref{p1} and \eqref{p2}.
This is because, if \eqref{p23} holds with strict inequality for any $t$,
we can always decrease the value of the related variable $q_t$ to reduce the total energy consumption
until \eqref{p23} is satisfied with the equality \cite{zyong}.

Problem \eqref{p2} is still non-convex, since it still involves two non-convex constraints, i.e., \eqref{p22} and \eqref{p23}.
Constraint \eqref{p23} can be tackled with the SCA method \cite{sca18}
by evaluating the global lower bound of \eqref{p23} at a given local point.
In particular, the left-hand side (LHS) of \eqref{p23} is a convex function of $q_t$, and the RHS is a jointly convex
function of $q_t$ and $(x_t,x_{t-1}, y_t, y_{t-1})$.
Since the first-order Taylor expansion serves as the global lower bound of a convex function \cite{Boyd},
we can obtain the following lower bound for the RHS of \eqref{p23}:
\begin{equation}\label{eq.q}
\begin{aligned}
&q_t^2 + \frac{(x_t - x_{t-1})^2 + (y_t - y_{t-1})^2}{\hat v_0^2} \geq q_t^{(\ell)2} + 2q_t^{(\ell)}(q_t - q_t^{(\ell)})  \\
&+\frac{2}{\hat v_0^2}[(x_t^{(\ell)} - x_{t-1}^{(\ell)})(x_t - x_{t-1})+(y_t^{(\ell)} - y_{t-1}^{(\ell)})(y_t - y_{t-1})]\\
&- \frac{1}{\hat v_0^2}[(x_t^{(\ell)} - x_{t-1}^{(\ell)})^2 + (y_t^{(\ell)} - y_{t-1}^{(\ell)})^2],
\end{aligned}
\end{equation}
where $q_t^{(\ell)}$, $x_t^{(\ell)}$, and $y_t^{(\ell)}$ are the respective values of the variables in the $\ell$-th iteration
of the SCA method.

We provide a lower bound for the concave function of the harvested solar power $P_s^t(z_t)$
by approximating it with a simple linear function, as given by
\begin{equation}\label{eq.solar3}
\hat P_s^t(z_t):= c_1 z_t + c_2,
\end{equation}
where $c_1$ and $c_2$ are two scaling coefficients depending on the parameters of energy harvesting.
We take the simple linear approximation of \eqref{eq.solar3} to $P_s^t(z_t)$,
rather than the first-order Taylor expansion as done in \eqref{eq.q}, since the
first-order Taylor expansion provides an upper bound for a concave function, not a lower bound.
Under our system and parameter settings, 
the approximation \eqref{eq.solar3} serves as a tight lower bound for $P_s^t(z_t)$,
as will be numerically verified in Section \ref{sec.sim}.

With the above mathematic manipulations and by letting $P_h:= \sum_t \tilde P_h^t$, $P_v:= \sum_t P_v^t$, and $f:= \sum_t f_t$,
problem \eqref{p2} can be transformed to
\begin{subequations}\label{p3}
\begin{align}
&\min_{\{x_t, y_t, z_t, q_t, \forall t \}} (P_h + P_v - f) \label{p31}\\
&\text {s.t.} ~\sum_{n=1}^t (\tilde P_h^n + P_v^n ) \leq \sum_{n=1}^t \hat P_s^n(z_n) + \eta_0 E_0/ \delta, ~\forall t, \label{p32} \\
&\frac{1}{q_t^2} \leq \frac{2}{\hat v_0^2}[(x_t^{(\ell)} - x_{t-1}^{(\ell)})(x_t - x_{t-1})+(y_t^{(\ell)} - y_{t-1}^{(\ell)})(y_t - y_{t-1})] \notag \\
&\qquad~ -\frac{1}{\hat v_0^2}[(x_t^{(\ell)} - x_{t-1}^{(\ell)})^2 + (y_t^{(\ell)} - y_{t-1}^{(\ell)})^2]  \notag \\
&\qquad~ +q_t^{(\ell)2} + 2q_t^{(\ell)}(q_t - q_t^{(\ell)}),~\forall t, \label{p33}\\
&q_t \geq 0, ~\forall t \label{p34}, \\ 
&\eqref{eq.mob1} - \eqref{eq.mob4}, \eqref{p14}~\text{and}~\eqref{p13}. \notag
\end{align}
\end{subequations}
Given the global lower bound in \eqref{eq.q} (and, in turn, the global upper bound of $P_h^t$)
and the lower bound in \eqref{eq.solar3},
constraint \eqref{p32} tightens the original constraint \eqref{p12}.
When the constraints of \eqref{p3} are satisfied, the constraints of the original problem \eqref{p1} are satisfied;
not the other way around.
Therefore, the feasible solution region of \eqref{p3} is a subset of the feasible solution region of \eqref{p1},
and the optimal value of \eqref{p3} draws an upper limit to the optimal value of \eqref{p1}.

All the constraints of \eqref{p3} are now convex.
Yet, the problem is still not convex due to the non-convex part of the objective function \eqref{p31}, i.e., $-f$,
which prevents a direct use of any standard convex solvers.
We propose to use the PDCAE to convexify \eqref{p31} and obtain a low-complexity (suboptimal) solution
for \eqref{p3} and, in turn, the original problem \eqref{p1}.

\subsection{PDCAE-based Solution}

The objective function of \eqref{p3} has the same form as the classic DC problem:
\begin{equation}\label{eq.dc}
\min_{x \in {\mathbb R}^n} G_0(x) + G_1(x) - G_2(x),
\end{equation} 
where $G_0$ is a smooth convex function with a Lipschitz continuous gradient and a Lipschitz continuity modulus;
$G_1$ is a proper closed convex function;%
\footnote{For an extended real-valued function $h: {\mathbb R}^n \rightarrow (-\infty, \infty)$,
whose domain is denoted by dom $h=\{x \in {\mathbb R}^n: h(x)< \infty \}$,
$h$ is a proper function if it never equals $-\infty$ and dom $h \ne \varnothing$.
Furthermore, $h$ is a proper closed function if it is lower semicontinuous \cite{wen18}.}
and $G_2$ is a continuous convex function.
In this sense, \eqref{p3} can be solved with the proximal DC algorithm~\cite{gotoh18},
which however, is slow to converge.

Developed in~\cite{wen18}, PDCAE accelerates the proximal DC algorithm with an extrapolation technique.
The extrapolation adds \emph{momentum} terms that depend on the solutions in the previous iterations
to update the variables during the current iteration\cite{nest07}.
The extrapolation has been widely used to speed up the proximal gradient algorithm and its variants
for convex optimization problems~\cite{nest13}.


We rewrite \eqref{p3} into the standard form of \eqref{eq.dc}.
The mapping between \eqref{p3} and \eqref{eq.dc} is as follows:
$P_v := G_0$, $P_h := G_1$, and $f := G_2$,
where $\mathbf X := x$ and $\boldsymbol \lambda_{(\ell)} := \theta_{(\ell)}$.
Since $f(\cdot)$ is a continuously differentiable function, we can obtain its gradient directly.
Further let $\mathbf X_t := [x_t, y_t, z_t, q_t], \forall t$,
and $\mathbf X := \{\mathbf X_t, \forall t \}$ collect the optimization variables at time slot $t$ and over the entire time horizon, respectively;
and $\cal A$ denote the feasible solution region of Problem \eqref{p3}.

The PDCAE-based solver of Problem \eqref{p3} is summarized in Algorithm~\ref{algo.proposed}, where
$\boldsymbol \lambda_t^{(\ell)}:= [\lambda_{xt}^{(\ell)}, \lambda_{yt}^{(\ell)}, \lambda_{zt}^{(\ell)}, \lambda_{qt}^{(\ell)}], \forall t$,
and $\boldsymbol \lambda_{(\ell)}:=\{ \boldsymbol \lambda_t^{(\ell)}, \forall t \}$.
$\lambda_{xt}^{(\ell)}$ denotes the extrapolation of the $x$-coordinate of the monitor at time slot $t$ in the $\ell$-th iteration,
\begin{equation}
\lambda_{xt}^{(\ell)} = x_t^{(\ell)} + \beta_{(\ell)}(x_t^{(\ell)} - x_t^{(\ell-1)}).
\end{equation}
Similarly, $\lambda_{yt}^{(\ell)}$ and $\lambda_{zt}^{(\ell)}$ denote the extrapolations of the monitor's $y$- and $z$-coordinates,
respectively; and $\lambda_{qt}^{(\ell)}$ the extrapolation of $q_t$.

The specific expressions in Step 5 of Algorithm 1 are given as follows.
\begin{subequations}\label{eq.gradient}
\begin{align}
&\nabla P_v^t(\boldsymbol \lambda_t^{(\ell)}) := \nabla P_v^t = \left[ \frac{W}{\delta}, -\frac{W}{\delta} \right], \\
&\nabla f_t =  \big[ 2\mu_1(x_t - a_t), 2\mu_1(y_t - b_t), 2\mu_2(z_t - z_{t-1}) \notag \\ 
&\qquad \quad \;  -2\mu_2(z_t - z_{t-1}) \big], \\
&\langle \nabla P_v^t(\boldsymbol \lambda_t^{(\ell)}) - \nabla f_t(\mathbf X_t^{(\ell)}), \mathbf X_t \rangle
= -2\mu_1(x_t^{(\ell)} - a_t)x_t \notag \\
&-2\mu_1(y_t^{(\ell)} - b_t)y_t +\left( W/ \delta - 2\mu_2 (z_t^{(\ell)} - z_{t-1}^{(\ell)}) \right)(z_t-z_{t-1}).
\end{align}
\end{subequations}
Recall that $P_v:= \sum_t P_v^t$ and $f:= \sum_t f_t$ in \eqref{p3}.
We have $\nabla P_v = \sum_t \nabla P_v^t$, and $\nabla f= \sum_t \nabla f_t$ in Step 5.

A popular way to select and update the extrapolation parameters $\{\beta_{(\ell)}\}$ is to set $\bar \beta_{(-1)} =\bar \beta_{(0)} =1$
and follow the recursive equations~\cite{wen18}
\begin{subequations}
\begin{align}
&\beta_{(\ell)} = (\bar \beta_{(\ell-1)} -1)/ \bar \beta_{(\ell)}; \\
&\bar \beta_{(\ell+1)} = \frac{1}{2}\left(1+\sqrt{1+4 \bar \beta_{(\ell)}^2}~ \right).
\end{align}
\end{subequations}
We choose a fixed number $\bar L$, and reset $\bar \beta_{({\bar L}-1)} =\bar \beta_{({\bar L})} =1$ every $\bar L$ iterations.

Given the same structure between problems \eqref{p3} and \eqref{eq.dc} and by following the same lines as in \cite{wen18},
the proposed PDCAE-based algorithm can converge to a suboptimal solution for the original problem \eqref{p1}
within polynomial time.

\begin{algorithm}[t]
\caption{The PDCAE-based solution for problem \eqref{p3}}
\label{algo.proposed}
\begin{algorithmic}[1]
\State {\bf Initialization:} Input $\mathbf X_{(0)} \in {\cal A}$, $\beta_{(\ell)} \subseteq [0,1), \forall \ell$.
Set $\mathbf X_{(-1)} = \mathbf X_{(0)}$.
\For {$\ell$ = 0, 1, 2, ...}
\State Compute the gradients $\nabla f(\mathbf X_{(\ell)})$.
\State Apply the extrapolation technique to the previous two iterations: $\boldsymbol \lambda_{(\ell)} = \mathbf X_{(\ell)} +
\beta_{(\ell)}(\mathbf X_{(\ell)} - \mathbf X_{(\ell-1)})$.
\State Update the optimization variables $\mathbf X_{(\ell+1)} = \argmin_{\mathbf X \in {\cal A}}
\{\langle \nabla P_v(\boldsymbol \lambda_{(\ell)})- \nabla f(\mathbf X_{(\ell)}), \mathbf X \rangle
+\frac{M}{2} || \mathbf X - \boldsymbol \lambda_{(\ell)} ||^2 + P_h (\mathbf X)\}$ by the interior point method.
\State Update $\ell \leftarrow \ell+1$.
\EndFor
\end{algorithmic}
\end{algorithm}

\begin{algorithm}[t]
\caption{The overall algorithm for problem \eqref{p1}}
\label{algo.overall}
\begin{algorithmic}[1]
\State Approximate the harvested solar power $P_s^t(z_t)$ by \eqref{eq.solar3}.
\State {\bf Initialization:} Input $\mathbf X_{(0)} \in {\cal A}$, $\beta_{(\ell)} \subseteq [0,1), \forall \ell$.
Set $\mathbf X_{(-1)} = \mathbf X_{(0)}$.
\For {$\ell$ = 0, 1, 2, ...}
\State Introduce slack variable $q_t, \forall t=1,\cdots, N$ based on \eqref{mu} and approximate the propulsion power $P_h^t$ by \eqref{newpm} and \eqref{p23} with the SCA method.
\State Transform problem \eqref{p1} to problem \eqref{p3}.
\State Implement Steps 3 to 5 of Algorithm \ref{algo.proposed} to obtain a suboptimal solution.
\State Update $\ell \leftarrow \ell+1$.
\EndFor
\end{algorithmic}
\end{algorithm}

\subsection{Overview of the Proposed Algorithm}

Algorithm \ref{algo.overall} summarizes the proposed algorithm, which solves problem \eqref{p1}
by rewriting \eqref{p1} as \eqref{p3} given $\mathbf{X}_t^{(\ell)}=(x_t^{(\ell)}, y_t^{(\ell)},z_t^{(\ell)},q_t^{(\ell)}),\,\forall t=1,\cdots,N$,
and then running Algorithm \ref{algo.proposed} to solve \eqref{p3} and obtain $\mathbf{X}_t^{(\ell+1)}$.
Algorithm \ref{algo.overall} repeats these steps until convergence,
i.e., $\|\mathbf{X}_t^{(\ell+1)}-\mathbf{X}_t^{(\ell)}\|<\epsilon$, where $\epsilon$ is the required accuracy of convergence.  

Problem \eqref{p1} is challenging due to the non-convex propulsion power function \eqref{speed},
harvested solar power function \eqref{eq.solar2}, and the concave part $-f_t$ in the objective function.
No known technique is able to obtain the globally or locally optimal solution to problem \eqref{p1}.
As proved in \cite{wen18}, the objective of \eqref{p31} is non-increasing and lower bounded,
and the proposed Algorithm \ref{algo.overall} can converge efficiently to a suboptimal solution.

{\blue
The computational complexity of Algorithm 2 is dominated by Step 6, and the step is further comprised of Steps 3 to 5 of Algorithm 1.
In Algorithm 1, Step 3 involves computing the gradients with solutions obtained at the previous iteration,
and Step 4 applies the extrapolation technique to the previous two iterations.
Both of their complexities are $\mathcal{O}(N)$.
In Algorithm 1, Step 5 updates the optimization variables with the PDCAE-based method.
Since the PDCAE deals with a convex problem after convexification, transformation and extrapolation,
the computational complexity of Step 5 is primarily accounted for by the interior point method,
which is $\mathcal{O}(N^{3.5})$ per iteration.
Taking all the steps into account, the total complexity of Algorithm 2 is $\mathcal{O}(N^{3.5})$ per iteration.
With the convergence accuracy $\epsilon$, the overall computational complexity of Algorithm~2 is
$\mathcal{O}(N^{3.5}\log\frac{1}{\epsilon})$,
where $\mathcal{O}(\log\frac{1}{\epsilon})$ gives the number of iterations required for convergence.

We note that Algorithm \ref{algo.overall} is suboptimal, due to the fact that the future trajectory of the target
is hard to predict in practice and, more importantly, the considered problem \eqref{p1} is non-convex.
Nevertheless, the algorithm is convergent and stable with a polynomial time-complexity, as discussed above;
and can be used online to refine the trajectory on-the-fly, as will be described in Section \ref{sec.extend}.

Also note that Algorithm \ref{algo.overall} is susceptible to the weather condition.
For example, the harvested solar energy can be substantially reduced if there are thick clouds,
which can have impact on the trajectory and heading of the monitor.
Nevertheless, the algorithm can deal with this situation from the following two aspects.
The first aspect is that Algorithm~\ref{algo.overall} plans the trajectory and heading of the monitor,
adapting to the harvested solar energy.
The impact of the cloud on the solar power harvesting has been explicitly considered in the algorithm.
In particular, the coefficient of atmospheric extinction $\alpha$ in \eqref{eq.solar} indicates that the intensity of solar beam decreases
in the presence of dense clouds.
The second aspect is that the monitor is equipped with a rechargeable battery,
which can supplement the harvested solar power when needed.
In the presence of thick clouds, the monitor can withdraw energy from the battery,
as the output of the battery is scheduled to power the monitor across the time slots, as captured in \eqref{p12}.
When the sky is clear, the battery can be recharged by any surplus solar energy.
This is because the battery needs to maintain at least a preconfigured minimum level for safety and emergency,
as also specified in \eqref{p12}.
The battery can be either discharged or recharged, depending on the abundance of the solar power
and the complexity of the monitoring mission.
}

\section{Extension to Online Implementation}\label{sec.extend}
In practice, it can be difficult for the monitor to predict precisely the entire trajectory of the target beforehand.
Instead, the monitor estimates the future waypoints of the target,
and adjusts its own trajectory accordingly.
The proposed algorithm provides a modularized solution which can be a critical building block of online implementation.
In this section, we generalize the proposed approach to online operations,
where the predictions of future waypoints are based on
the monitor's past observation and used to plan progressively the monitor's trajectory.

\begin{figure}[t]
\centering
\includegraphics[width=0.5\textwidth]{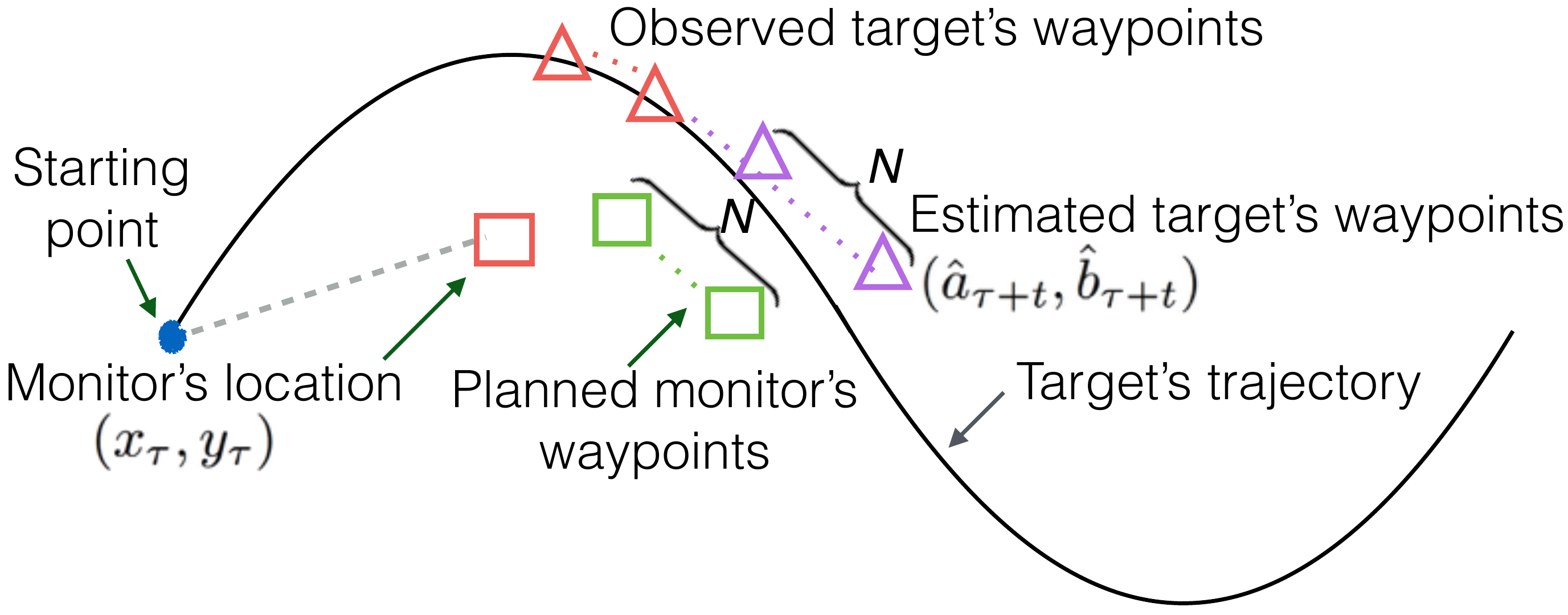}
\caption{An illustration of the online scheme in the 2D space.}
\label{online}
\end{figure}

{\blue
The online approach is an extension of the offline Algorithm~\ref{algo.proposed},
where Algorithm \ref{algo.proposed} serves as the dynamic method for the online control of the monitoring UAV.
Specifically, at any instant $\tau$, the monitor predicts the target's trajectory over the next $N$ instants.
Following the model-predictive control (MPC)~\cite{mpc13},
the monitor can run Algorithm \ref{algo.proposed} on-the-fly to plan its trajectory from its current location for the next $N$ instants (or steps).
Only the first step of the $N$-step trajectory is implemented at instant $(\tau+1)$ though.
Then, the monitor predicts the target's trajectory over the upcoming $N$ instants.
Algorithm~\ref{algo.proposed} is run again starting from the new current location of the monitor, yielding a new control based on the new prediction.
The prediction horizon of $N$ instants (or steps) keeps shifting forwards (by one at every instant).
The details are provided in the following.
}

We suppose that the monitor keeps observing the locations (i.e., the coordinates) and headings of the target.
At any (current) time slot $\tau$, the monitor predicts the waypoints of the target for the upcoming $N$ slots, i.e.,
$(\hat a_{\tau+t}, \hat b_{\tau+t}), t =1,\ldots, N$ based on the past observation of the target.
The monitor's current waypoint is $(x_{\tau}, y_{\tau}, z_{\tau})$.
As illustrated in Fig.~\ref{online}, the monitor plans its trajectory for the upcoming $N$ slots (i.e., slots $\tau+1, \cdots, \tau+N$),
by solving the following problem at time slot $\tau$:
\begin{subequations}\label{p.onmode1}
\begin{align}
&\min_{\{x_t, y_t, z_t, \forall t =1, \cdots, N\}} \sum_{n=\tau+1}^{\tau+N} (P_h^n+P_v^n - f_n) \label{p.onmode11}\\
&\text {s.t.} ~ \sum_{n=1}^{\tau+t} (P_h^n + P_v^n ) \leq \sum_{n=1}^{\tau+t} P_s^n(z_n)+ \eta_0 E_0/ \delta 
, \forall t, \label{p.onmode12}\\
&(x_{\tau+t}-x_{\tau+t-1})^2+(y_{\tau+t}-y_{\tau+t-1})^2 \leq (V_{hm} \delta)^2, ~\forall t, \label{p.onmode13} \\
&| z_{\tau+t}- z_{\tau+t-1} - (z_{\tau+t-1} - z_{\tau+t-2}) | \le V_{vm} \delta, ~\forall t \ge 2, \label{p.onmode14} \\
&(x_{\tau+t}, y_{\tau+t}, z_{\tau+t}) \in \hat{\cal{F}}, \label{p.onmode15}
\end{align}
\end{subequations}
where $\hat{\cal{F}}$ is the new FFR, as given by
\begin{align}\label{eq.ffr2}
\hat{\cal{F}} := &\{(x_{\tau+t}, y_{\tau+t}, z_{\tau+t})|H < z_l \le z_{\tau+t}, \notag \\
&~ x_{\tau+t} \le \hat{a}_{\tau+t}, ~y_{\tau+t} \le \hat{b}_{\tau+t}, \notag \\
&~ (x_{\tau+t}-\hat{a}_{\tau+t})^2+(y_{\tau+t}-\hat{b}_{\tau+t})^2 + (z_{\tau+t}-H)^2 \notag \\
&~\le D^2, ~\forall t \},
\end{align}
and \eqref{p.onmode12} guarantees that the monitor can always maintain a minimum energy reserve of $(1-\eta_0)E_0$ Joules in the battery
at any time slot.

{\blue
As with problem \eqref{p1}, the non-convex propulsion and solar power functions in \eqref{p.onmode1} can be tightened
linearly and iteratively with the SCA method, resulting in the following DC problem:
\begin{subequations}\label{p.onmode2}
\begin{align}
&\min_{\{x_t, y_t, z_t, \forall t =1, \cdots, N\}} \sum_{n=\tau+1}^{\tau+N} (\tilde P_h^n+P_v^n - f_n) \label{p.onmode21}\\
&\text {s.t.} ~ \sum_{n=1}^{\tau+t} (\tilde P_h^n + P_v^n ) \leq \sum_{n=1}^{\tau+t} \hat P_s^n(z_n)+ \eta_0 E_0/ \delta 
, \forall t, \label{p.onmode22}\\
&\frac{1}{q_t^2} \leq \frac{2}{\hat v_0^2}[(x_t^{(\ell)} - x_{t-1}^{(\ell)})(x_t - x_{t-1})+(y_t^{(\ell)} - y_{t-1}^{(\ell)})(y_t - y_{t-1})] \notag \\
&\qquad~ -\frac{1}{\hat v_0^2}[(x_t^{(\ell)} - x_{t-1}^{(\ell)})^2 + (y_t^{(\ell)} - y_{t-1}^{(\ell)})^2]  \notag \\
&\qquad~ +q_t^{(\ell)2} + 2q_t^{(\ell)}(q_t - q_t^{(\ell)}),~\forall t, \label{p.onmode23}\\
&q_t \geq 0, ~\forall t \label{p.onmode24}, \\ 
& \eqref{p.onmode13} - \eqref{p.onmode15}. \notag
\end{align}
\end{subequations}

Problem \eqref{p.onmode2} exibits the same structure as \eqref{p3}.
By replacing the actual waypoints of the target $(a_t, b_t)$ in Algorithm \ref{algo.proposed} with the estimated waypoints
$(\hat a_{\tau+t}, \hat b_{\tau+t})$, $t=1,\cdots,N$,
problem \eqref{p.onmode2} can be readily solved by using the PDCAE-based Algorithm \ref{algo.proposed}.
In this sense, the algorithm provides the so-called model which is based on regularly updated observations and predictions,
and can be solved in a structured manner.

Algorithm 3 summarizes the online implementation of the proposed approach by following the MPC framework.
Under the MPC framework, the monitor solves \eqref{p.onmode2} to produce a tentative trajectory of $N$ waypoints
for the upcoming $N$ time slots, i.e., $\tau+1,\cdots,\tau+N$, at every slot $\tau=1,2,\cdots$.
The monitor only takes the trajectory at the next time slot, i.e., $\tau+1$, and then solves \eqref{p.onmode2} again
to update the trajectory.}
By this means, the MPC can adapt quickly to new inputs and provide the control policy with consideration of future events,
achieving a fast, flexible and dynamic control.
The impact of prediction errors can be mitigated by exploiting the online adjustability of the MPC and incorporating the error bounds
into the control, thereby achieving an effective online control of the trajectory.
{\red Like Algorithm 2, the computational complexity of Algorithm 3 is also dominated by Step 5 in Algorithm 1.
Therefore, the complexity of Algorithm 3 is $\mathcal{O}(N^{3.5}\log\frac{1}{\epsilon})$.}

It is worth noting that the monitor can further enhance its disguise by regulating its horizontal heading to deviate from the heading of the target
at each time slot (in addition to keeping the horizontal distance and changing the altitude at each time slot,
as described in Sections \ref{sec.formulate} and \ref{sec.solution}).
This can be implemented by controlling the slope of the monitor's trajectory on the $(x,y)$-plane,
denoted by $\kappa_{\tau}$, at any time slot $\tau$.
The slope $\kappa_{\tau}$ can change every slot,
and deviate from the target's heading $\kappa_{ab,\tau}$ by at least a predefined nonnegative value $c_3>0$.

Considering the heading control of the monitor by increasing its slope, we have
\begin{equation}\label{eq.slope}
\kappa_{\tau}:=\frac{y_{\tau}-y_{\tau-1}}{x_{\tau}-x_{{\tau}-1}} \ge \frac{b_{\tau}-b_{\tau-1}}{a_{\tau}-a_{\tau-1}} + c_3, ~\forall \tau,
\end{equation}
which is a linear inequality, and can be imposed as an additional constraint of \eqref{p1} or \eqref{p.onmode1}.
Given the linearity of \eqref{eq.slope}, the convexification process of \eqref{p1} or \eqref{p.onmode1} is unaffected
and Algorithm 1 remains effective with further improved disguising,
as will be numerically validated in Section \ref{sec.sim}.

\begin{algorithm}[t]
\caption{\blue The proposed online algorithm for problem \eqref{p.onmode1}.}
\label{algo.online}
\begin{algorithmic}[1]
\State At time slot $\tau$, the monitor observes the current location of the target and predicts the target's next
$N$ waypoints.
\State Approximate the harvested solar power $P_s^t(z_t)$ by \eqref{eq.solar3}.
\State {\bf Initialization:} Set a feasible initial trajectory for the monitor, i.e.,
$\{ (x_{t}^{(0)}, y_{t}^{(0)}, z_{t}^{(0)}), t=1,\cdots,N\}$,
and initialize the auxiliary variables $\{\beta_{(0)}, q_{t}^{(0)}, \boldsymbol \lambda_{t}^{(0)}, t=1,\cdots,N\}$.
\For {$\ell$ = 0, 1, 2, ...}
\State Introduce slack variable $q_t, \forall t=1,\cdots, N$, approximate the propulsion power $P_h^t$
by \eqref{newpm} and \eqref{p23}, and tighten constraint \eqref{p.onmode12} with the SCA method.
\State Implement Steps 3 to 5 of Algorithm 1 to plan the monitor's trajectory for the upcoming $N$ slots.
\State Update $\ell \leftarrow \ell+1$.
\EndFor
\end{algorithmic}
\end{algorithm}


\section{Numerical Results}\label{sec.sim}
This section presents MATLAB simulation results to corroborate the merits of the proposed approach.
Three baseline schemes are compared by setting the values
of $\mu_1$ and/or $\mu_2$ to zero.
We also include the extended schemes of moving direction (i.e., heading) regulation (labeled as ``MDR'')
and MPC-based online optimization (labeled as ``Online'') described in Section \ref{sec.extend}.
Parameters concerning the monitoring UAV's power consumptions and harvested solar power are given in Table \ref{tab.pm}.
The total scheduling period is $T=30$ s with each time slot of $0.2$~s.
The maximum monitor-target 3D distance is $D=20$ m.
The constant controlling the slope of the monitor's heading is $c_3=1$.
The trajectory of the target is generated by $a_t = 10t$ and $b_t = 100 \sin(t/5)$, unless otherwise stated.
{\blue
We set the initial trajectory of the monitor to be the same as the (predicted) target's trajectory.
This initial trajectory is feasible, as it satisfies the UAV mobility and power consumption constraints.
}

\begin{table}[t]
\renewcommand{\arraystretch}{1.3}
\caption{The Parameters for UAV propulsion power, thrust power and harvested solar power \cite{zyong, kokh04}}
\begin{center}
\begin{tabular}{l | l} 
\hline
\text{Parameter}    &\text{Value} \\ \hline
\text{UAV weight}                    &4 kg \\ \hline
\text{Blade profile power and induced power, $P_0$, $P_1$}               &3.4 W, 118 W \\ \hline
\text{Rotor solidity and disc area, $s$, $A$}                       &0.03, 0.28 m$^2$\\ \hline
\text{Tip speed of the rotor blade, $U_{tip}$}     &60 m/s \\ \hline
\text{Mean rotor induce velocity, $v_0$ }     &5.4 m/s\\ \hline
\text{Atmospheric density and fuselage drag ratio, $\rho$, $d_f$}            &1.225 kg/m$^3$, 0.3 \\ \hline
\text{Maximum horizontal and vertical speed, $V_{hm}$, $V_{vm}$}        & 30 m/s, 8 m/s     \\ \hline
\text{Atmospheric extinction, $\alpha$}                        &0.8978 \\ \hline
\text{Power intensity of solar beams, $P_i$}               &1367 W/m$^2$ \\ \hline
\text{Efficiency and size of solar panel, $\eta$, $S$}                             &0.4, 0.5 m$^2$ \\ \hline
\text{Ratio of usable energy in the battery, $\eta_0$}                             &0.9 \\ \hline
\text{Initial altitude of the target and monitor, $H$, $z_0$}                             &100 m, 102 m \\ \hline
\text{Coefficients for solar power approximation, $c_1$, $c_2$}                             &0.0097, 165.83 \\ \hline
\end{tabular}
\end{center}
\label{tab.pm}
\end{table}

\begin{figure}[t]
\centering
\includegraphics[width=0.47\textwidth]{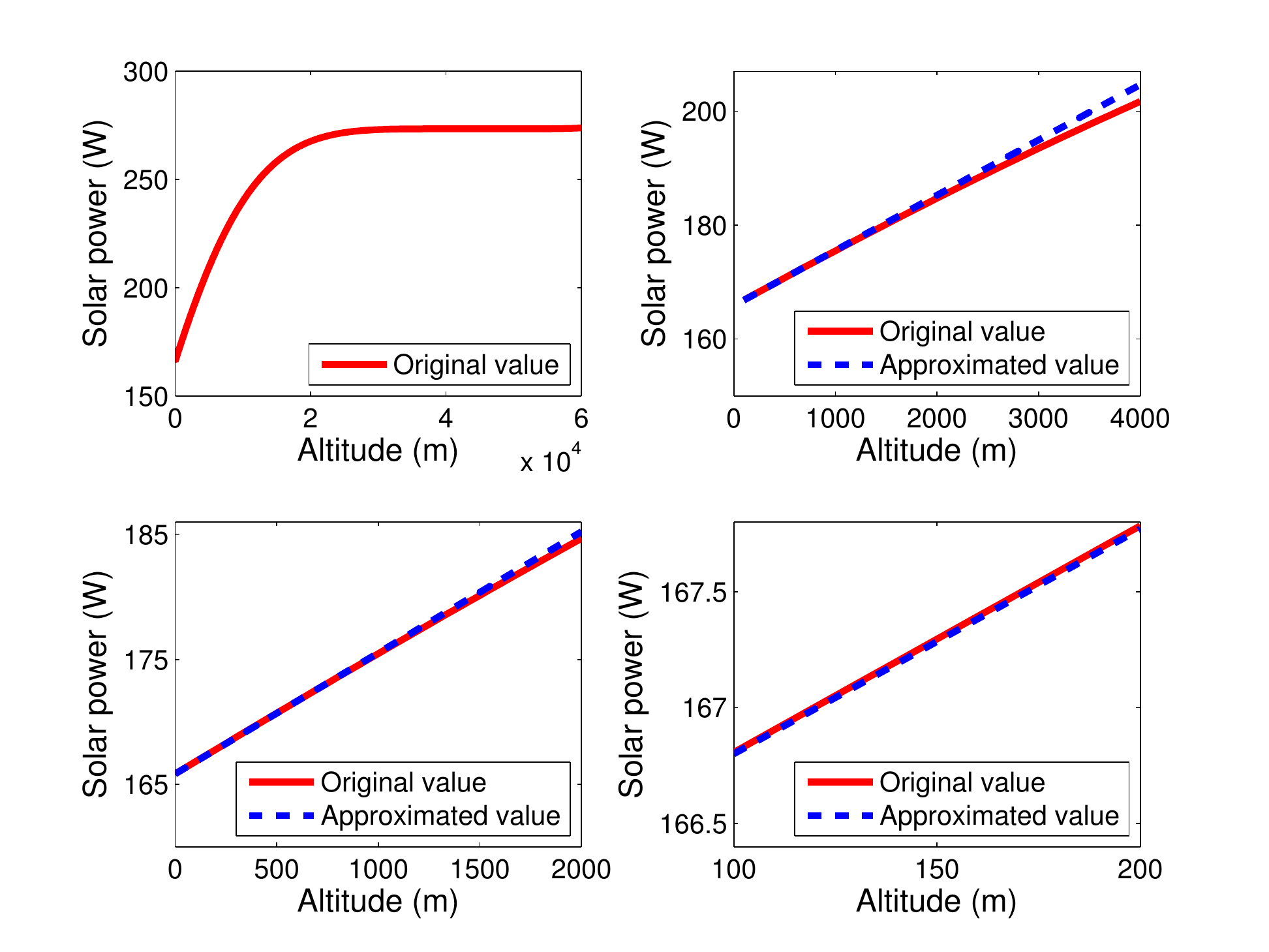}
\caption{Original value and approximation of solar power.}
\label{solar}
\end{figure}

We first show the tightness of the solar power approximation in \eqref{eq.solar3}.
The scaling coefficients of the linear function $c_1$ and $c_2$ are given in Table \ref{tab.pm}.
It can be seen from Fig. \ref{solar} that the linear approximation is tight for an altitude below $4000$ m
and almost overlaps with the original function when the altitude is below $2000$ m.
For the altitude typically ranging from $100$ m to $200$ m in our application scenario,
the approximation \eqref{eq.solar3} provides a very tight lower bound for the original value,
with a negligible approximation error of only $10^{-5}$ m.

\begin{figure}[t]
\centering
\includegraphics[width=0.47\textwidth]{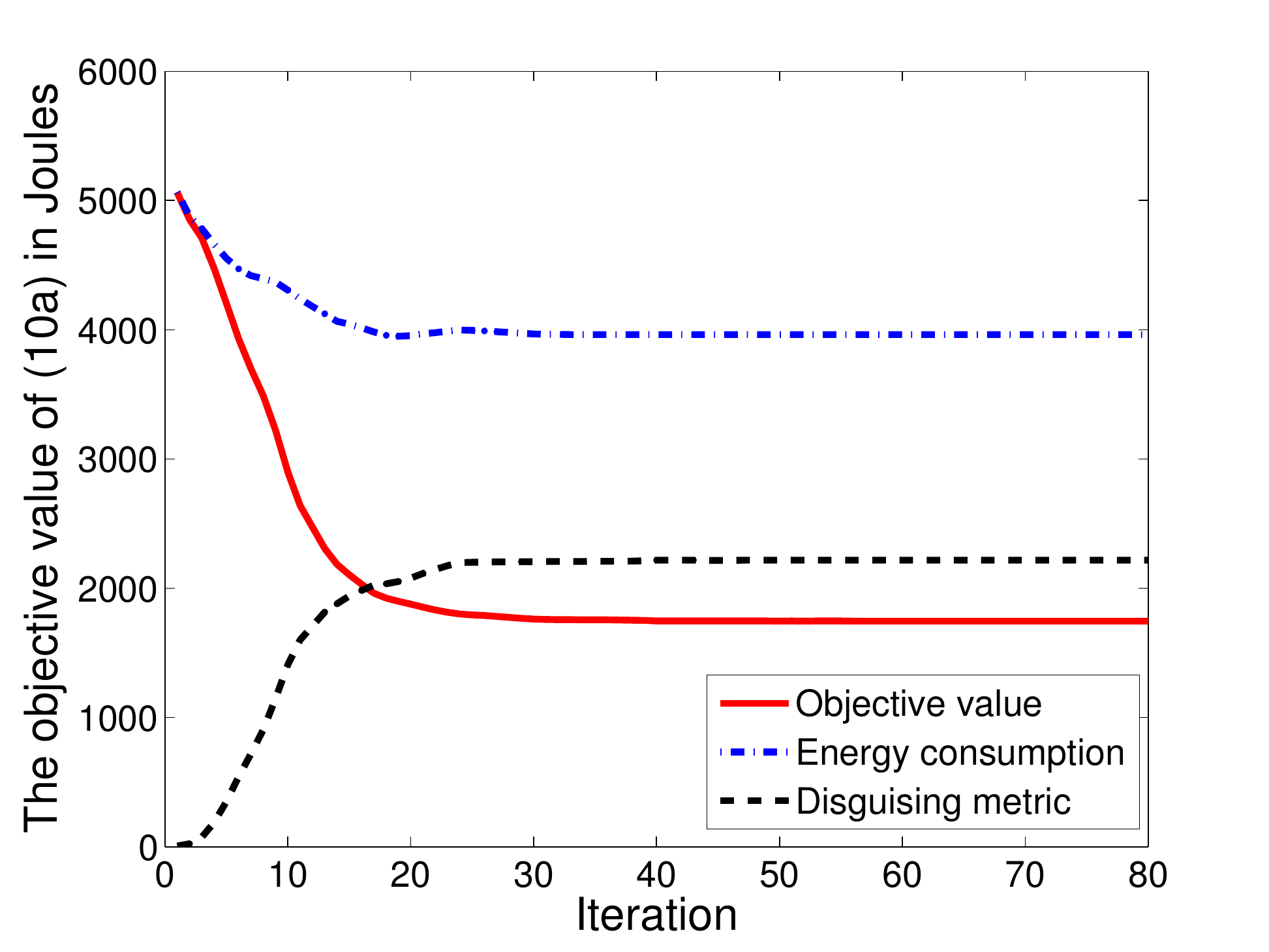}
\caption{Convergence of the original objective value in \eqref{p11}.}
\label{converge}
\end{figure}

\begin{figure}[t]
\centering
\includegraphics[width=0.47\textwidth]{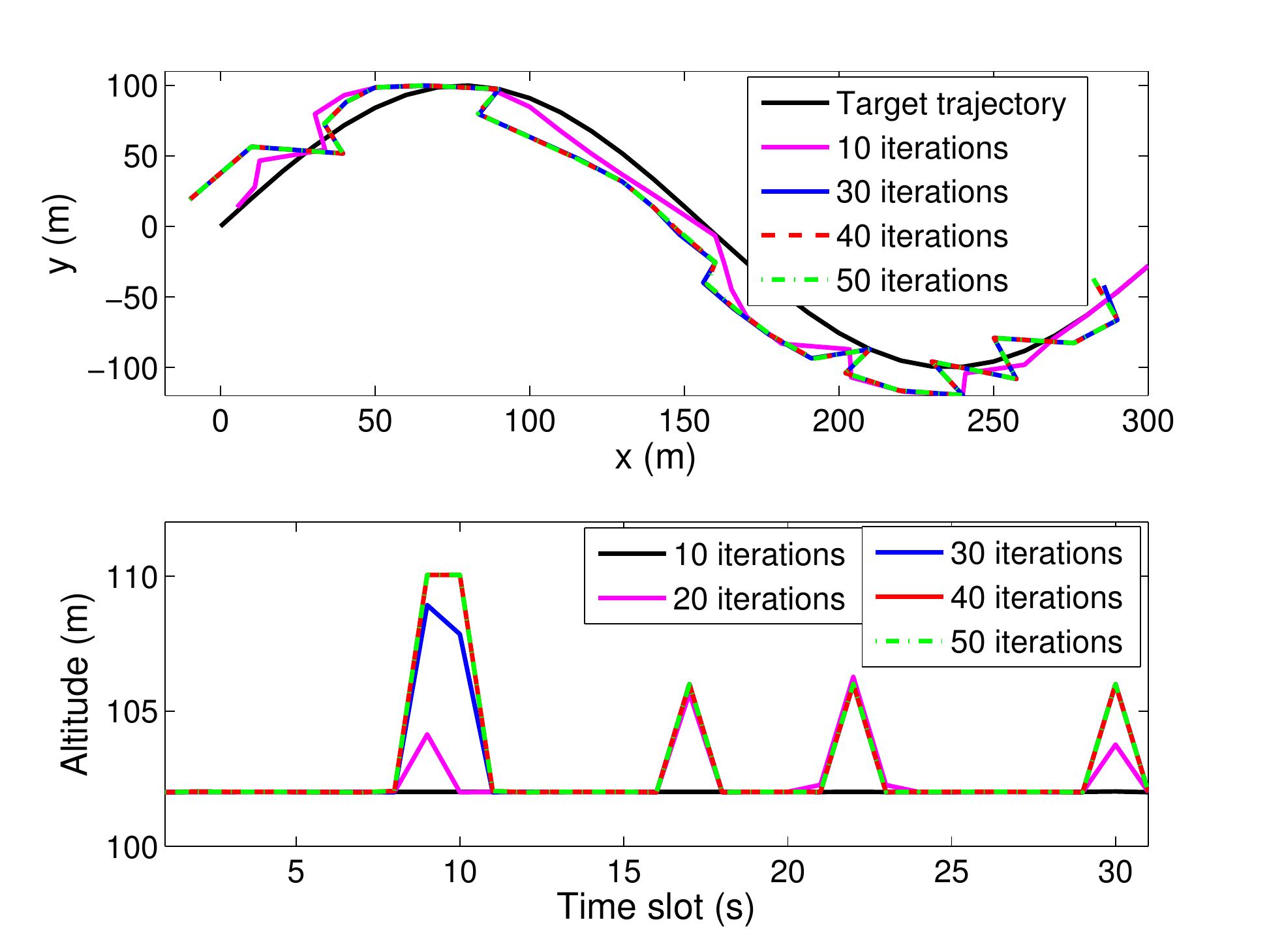}
\caption{Convergence of the monitor 3D trajectory.}
\label{route}
\end{figure}

\begin{figure}[t]
\centering
\includegraphics[width=0.47\textwidth]{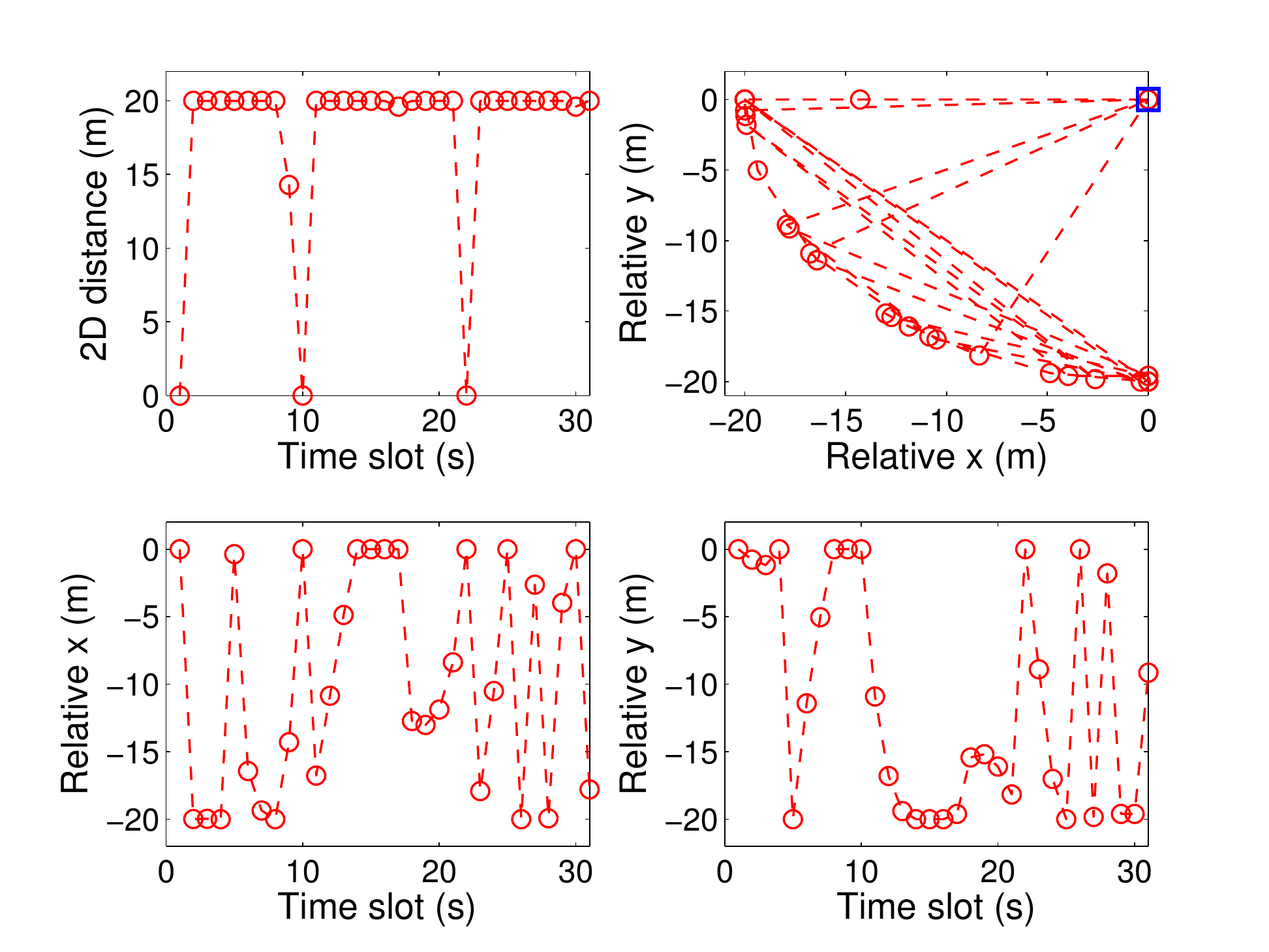}
\caption{Relative monitor 2D trajectory observed by the target.}
\label{relative}
\end{figure}

Fig. \ref{converge} shows the convergence speed of the proposed approach for problem \eqref{p1},
where $\mu_1=0.2$ and $\mu_2=0.1$.
The red solid line is the original objective value in problem \eqref{p1},
the blue dash-dot (non-increasing) line is the total energy consumption of the monitoring UAV $(P_h + P_v)\delta$ [cf. \eqref{p3}],
and the black dashed line is the total disguising performance $f$.
It is revealed in Fig. \ref{converge} that the objective value can converge fast within $60$ iterations.
The monitor 3D trajectory is depicted in Fig. \ref{route}.
It can be observed that both the horizontal trajectory and the altitude control converge within $60$ iterations.
Horizontally, the monitor tails the target and surveils in a distance.
Vertically, the monitor changes its altitude regularly, further deviating from the heading of the target.
Bounded by constraint \eqref{p14}, the monitor either ascends by the maximum height $\delta V_{vm}$ that it reaches in the first slot
and then stays at the altitude in the next slot,
or first ascends and then descends by $\delta V_{vm}/2$ in the first two slots.
This performance limitation averts a sharp ascend-then-descend action of the monitor at the full vertical speed,
protecting the generator and preventing a speed loss.

Fig.~\ref{relative} depicts the 2D monitor-target distance and the relative 2D trajectory of the monitor observed by the target
which is treated as a reference point and marked by a blue square.
The relative coordinates are $(x_t-a_t, y_t-b_t), \forall t$, and the target is at $(0,0)$.
It can be seen from Fig.~\ref{relative} that the monitor always flies in the southwest region to the target
at a distance of no more than $D~(=20)$ meters.
As observed by the target, the monitor exhibits random movement patterns.

\begin{figure}[t]
\centering
\includegraphics[width=0.47\textwidth]{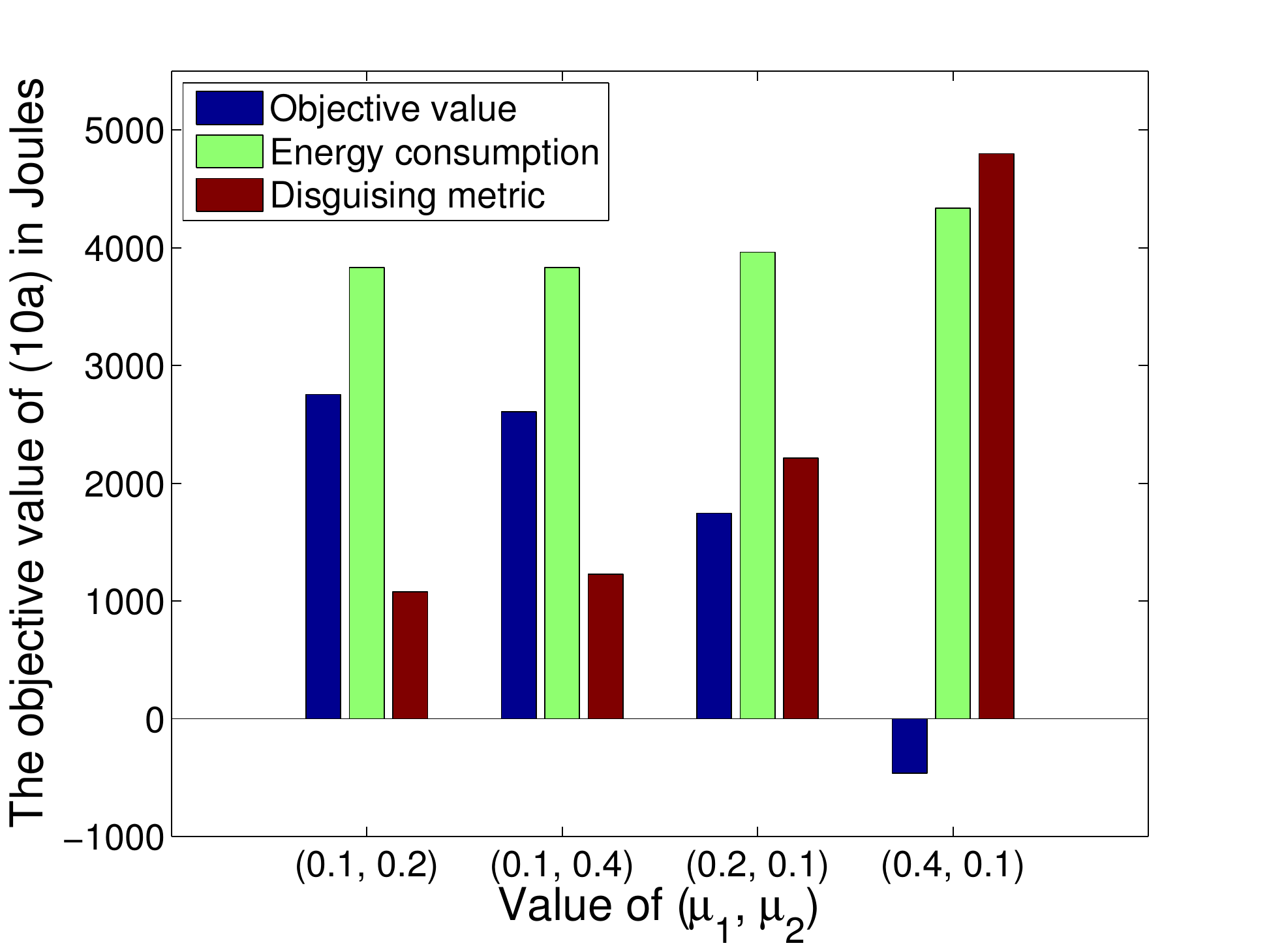}
\caption{Objective values under different values of ($\mu_1, \mu_2$).}
\label{objdiffmu}
\end{figure}

\begin{figure}[t]
\centering
\includegraphics[width=0.47\textwidth]{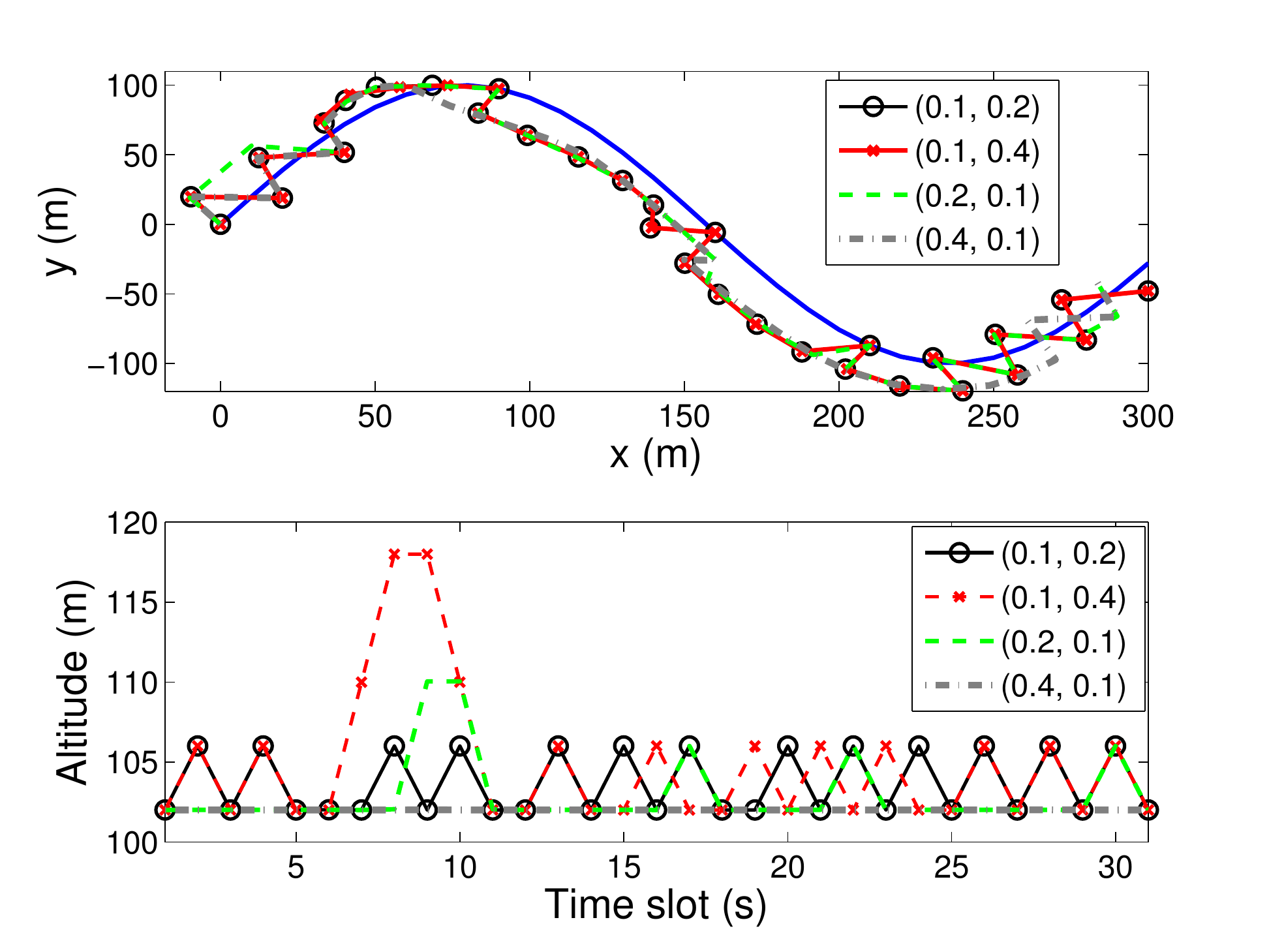}
\caption{Monitor 3D trajectory under different values of ($\mu_1, \mu_2$).}
\label{diffmu}
\end{figure}

\begin{figure}[t]
\centering
\includegraphics[width=0.47\textwidth]{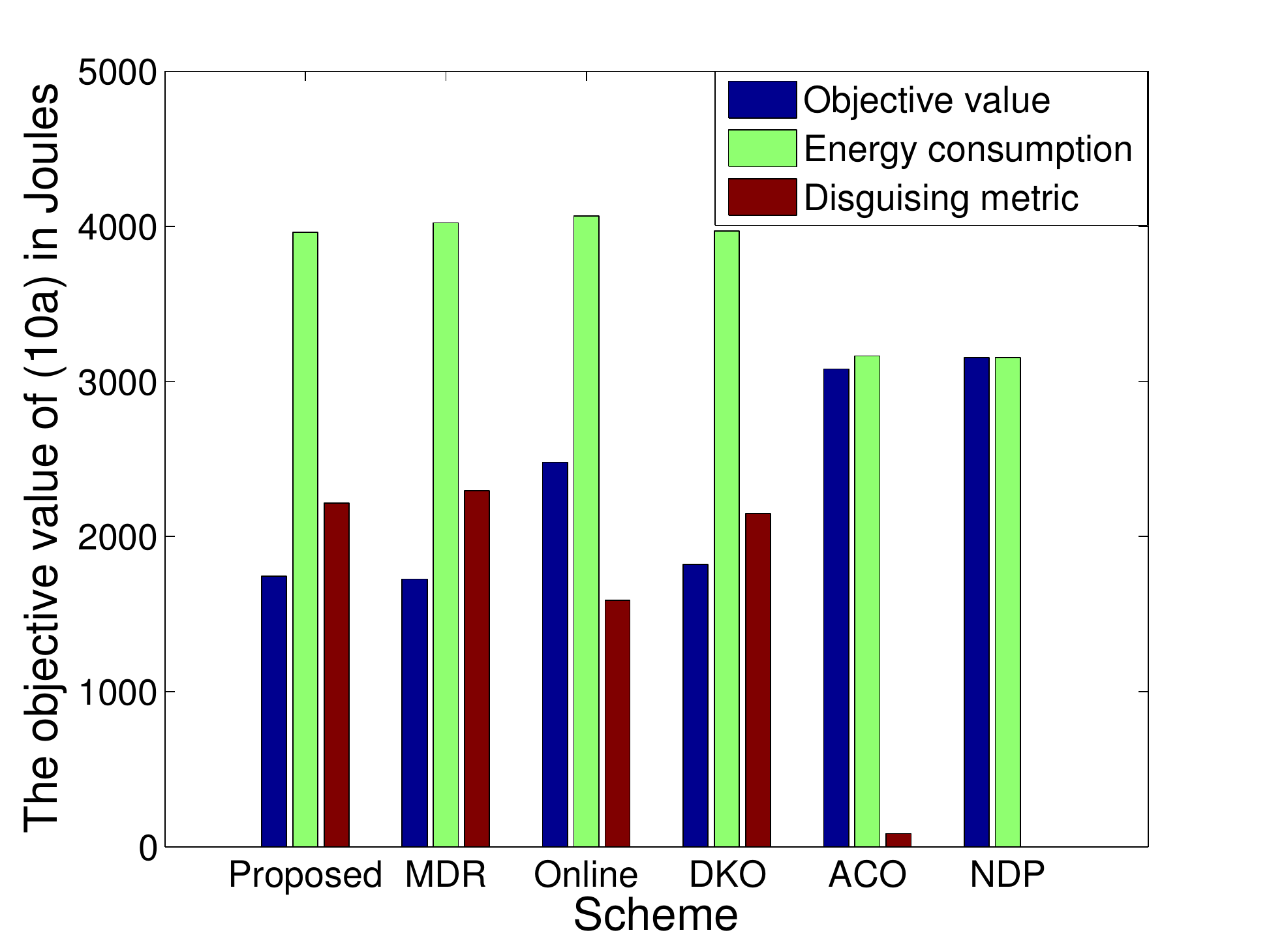}
\caption{Objective values under different schemes.}
\label{obj6}
\end{figure}

By changing the weights of the disguising performance $\mu_1$ and $\mu_2$, we observe their impact on the objective
and the trajectory of the monitoring UAV.
Fig. \ref{objdiffmu} gives the objective values of problem \eqref{p1} under different settings of $\mu_1$ and $\mu_2$.
By comparing the performances under $(0.1, 0.4)$ and $(0.4, 0.1)$, we can conclude that keeping the horizontal distance
is more efficient to reduce the objective value than changing the altitude.
Yet, increasing the weights of disguising cannot enhance the performance unlimitedly, as the monitor is also constrained
by its maximum speed and the 3D distance to the target.
Fig. \ref{diffmu} depicts the corresponding monitor 3D trajectory of each scenario,
where the blue solid line is the target trajectory.
It is shown in Fig. \ref{diffmu} that the altitude is more sensitive to the disguising performance than the horizontal distance.
When the weight of altitude control rises from $0.2$ to $0.4$, the altitude changes even more frequently,
which can cause the undesirable instability of the monitor and may trigger a speed loss.
For $(0.4, 0.1)$, the monitor always stays at the initial height, since the monitor-target horizontal distance is around $20$ m,
and there is no room for the monitor to rise or increase the vertical distance.

We also test the performances of the baseline schemes and the extended schemes.
The proposed scheme runs with $\mu_1=0.2$ and $\mu_2=0.1$, which is labeled as ``Proposed''.
The baseline schemes are the Proposed scheme with distance keeping only (labeled as ``DKO''), i.e., $\mu_1=0.2$ and $\mu_2=0$,
altitude changing only (labeled as ``ACO''), i.e., $\mu_1=0$ and $\mu_2=0.1$,
and no disguising performance (labeled as ``NDP''), i.e., $\mu_1=\mu_2=0$,
which essentially only minimizes the total energy consumption of the monitor.
The extended schemes are the MDR and the Online schemes.

\begin{figure}[t]
\centering
\includegraphics[width=0.47\textwidth]{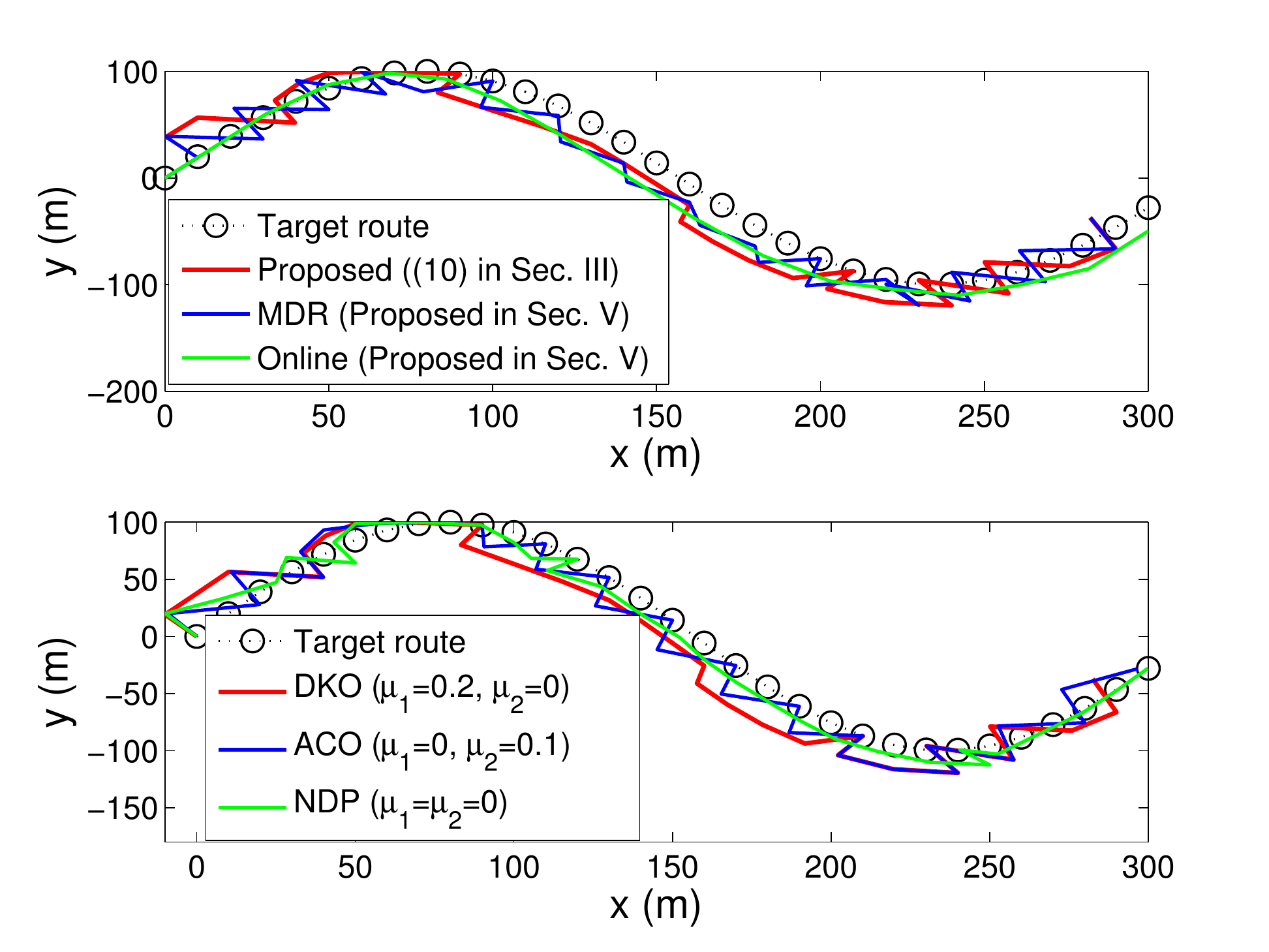}
\caption{Monitor horizontal trajectory under different schemes.}
\label{horizon6}
\end{figure}

\begin{figure}[t]
\centering
\includegraphics[width=0.47\textwidth]{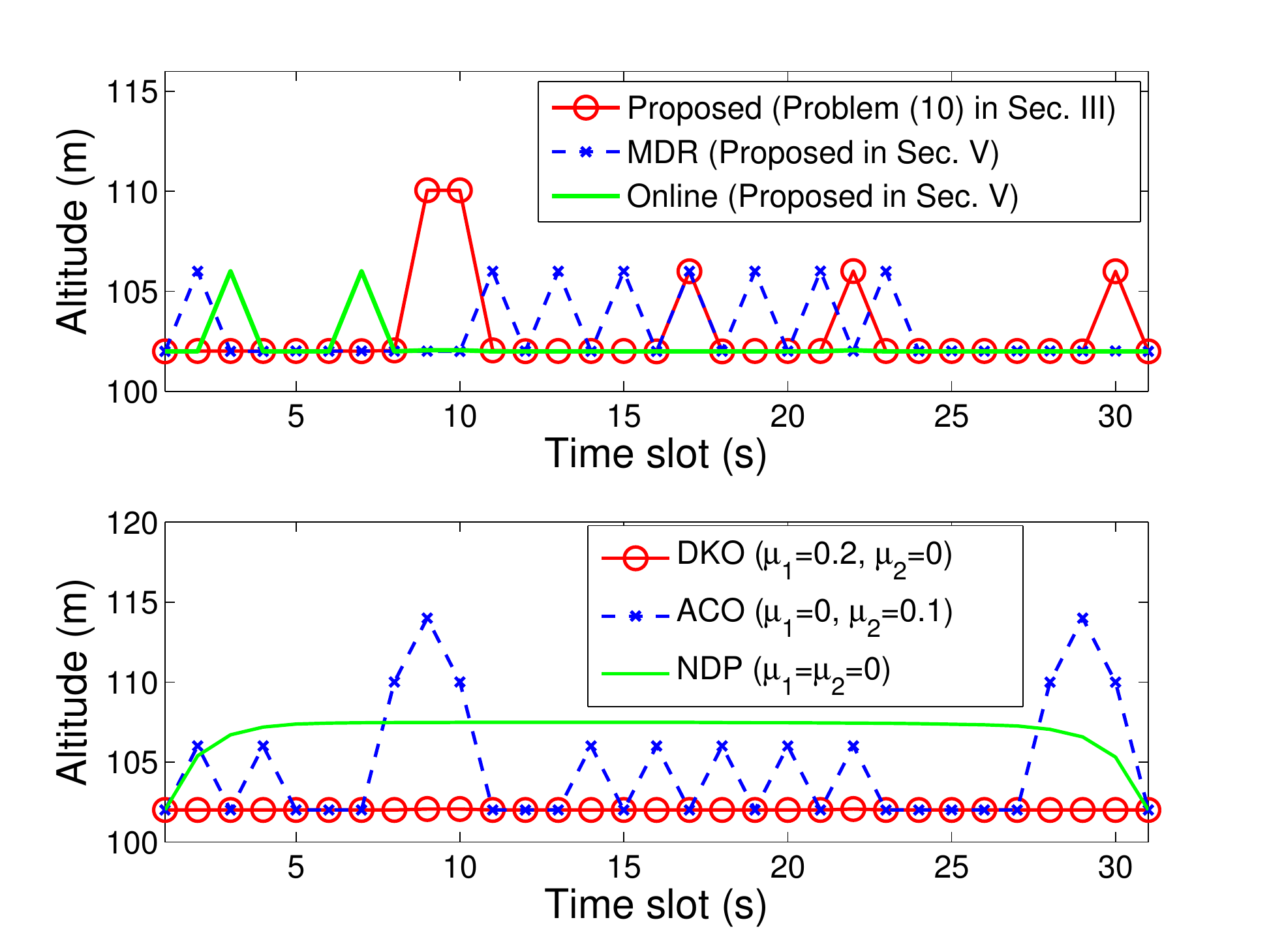}
\caption{Monitor altitude under different schemes.}
\label{height6}
\end{figure}

Fig. \ref{obj6} shows the objective values of the monitoring UAV, i.e., \eqref{p11}, under different optimization schemes.
Figs. \ref{horizon6} and \ref{height6} depict the corresponding horizontal and vertical trajectories of the monitor, respectively.
It is shown in Fig. \ref{obj6} that the Proposed scheme and the MDR scheme score the highest in terms of the disguising performance,
followed by the DKO scheme.
The Proposed scheme and the MDR scheme can thus significantly suppress the objective values.
By comparing the DKO scheme and the ACO scheme, it is also seen that keeping distance is much more effective
than changing altitude to enhance the disguising performance and reduce the objective value.
On the other hand, the NDP scheme yields the least energy consumption, as it is actually formulated that way.

\begin{figure}[t]
\centering
\includegraphics[width=0.47\textwidth]{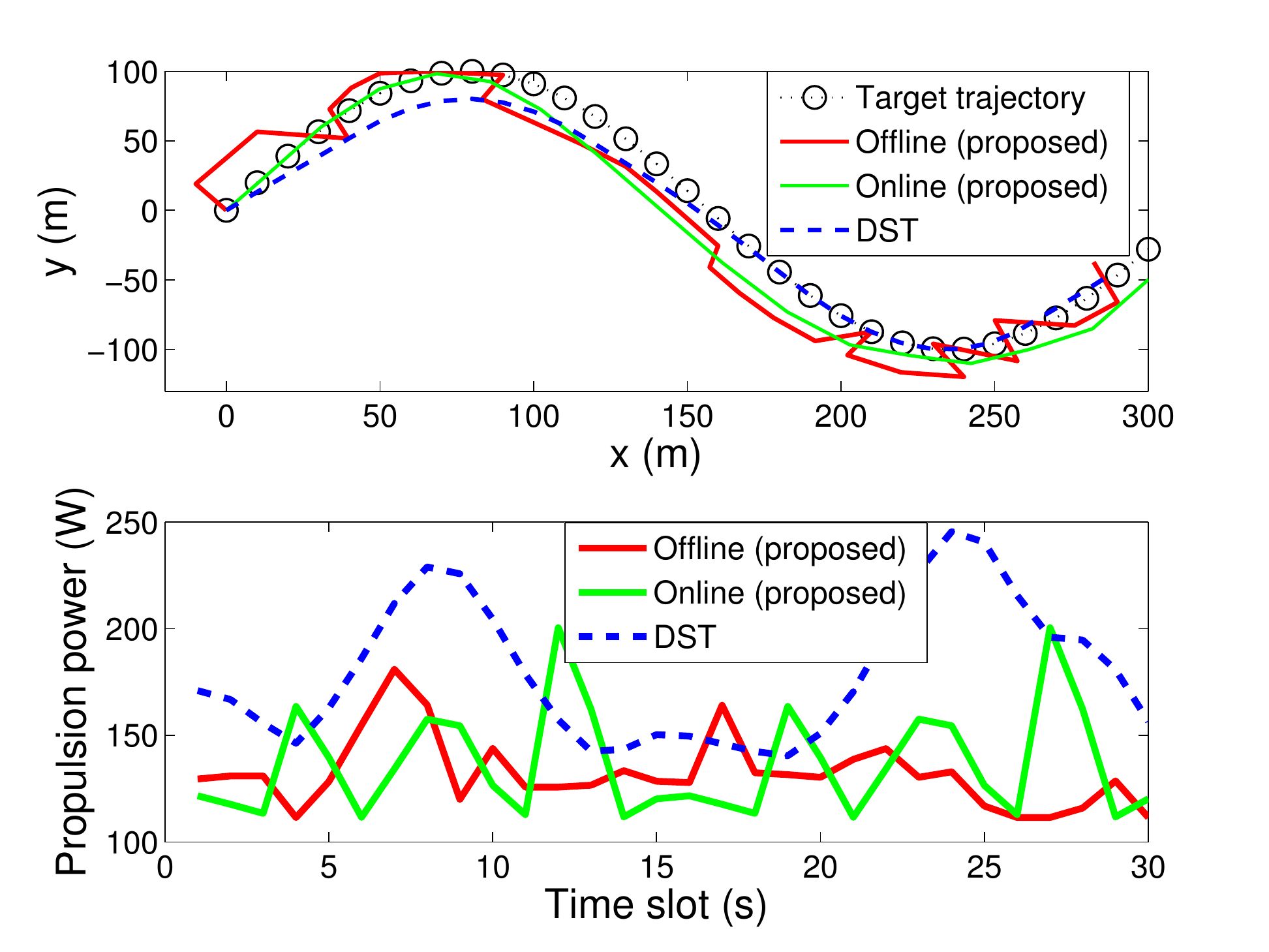}
\caption{Energy saving of the proposed schemes.}
\label{save}
\end{figure}

\begin{figure}[t]
\centering
\includegraphics[width=0.47\textwidth]{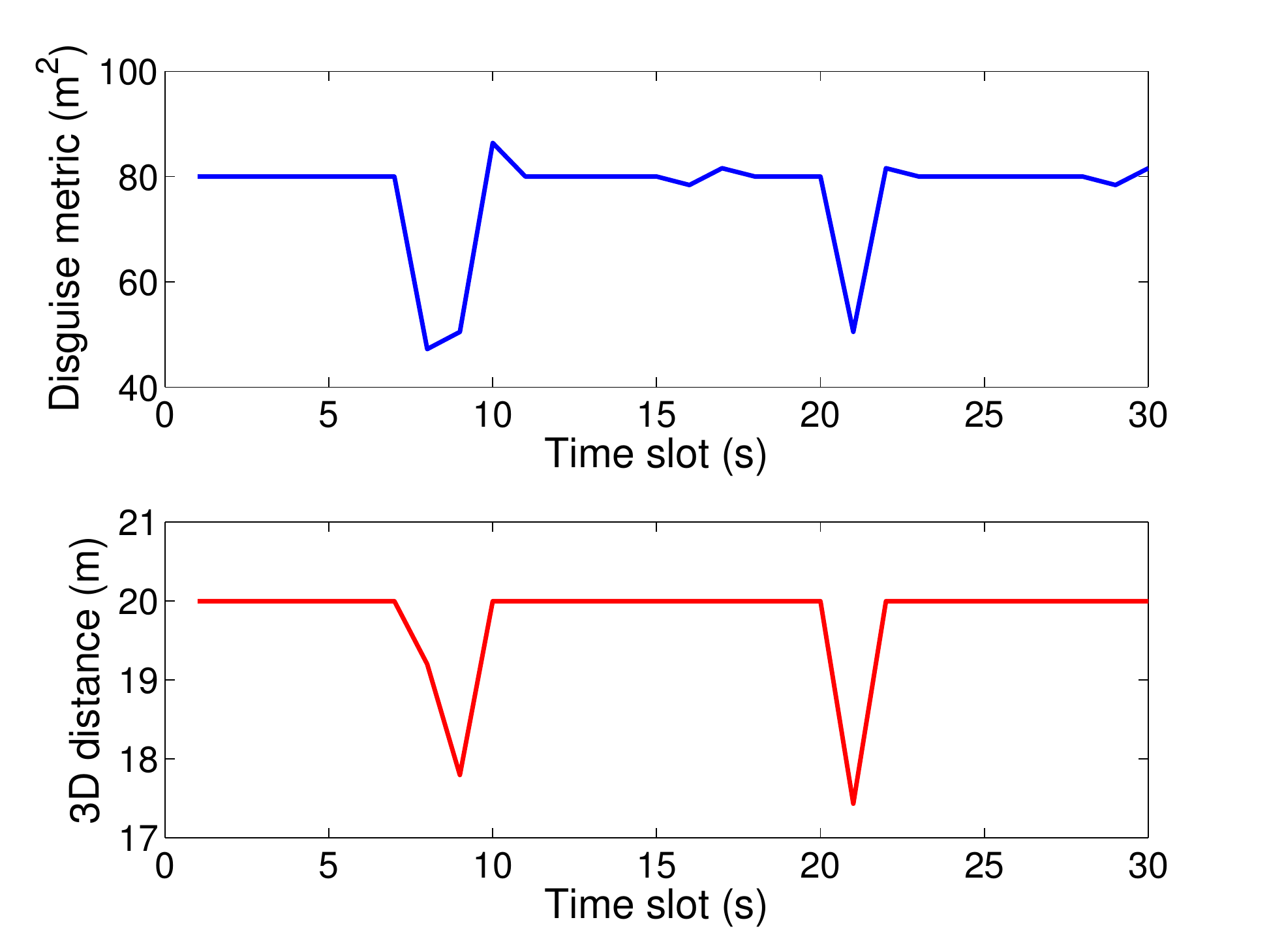}
\caption{Disguising metric and 3D monitor-target distance.}
\label{covdis}
\end{figure}


\begin{figure}[t]
\centering
\includegraphics[width=0.47\textwidth]{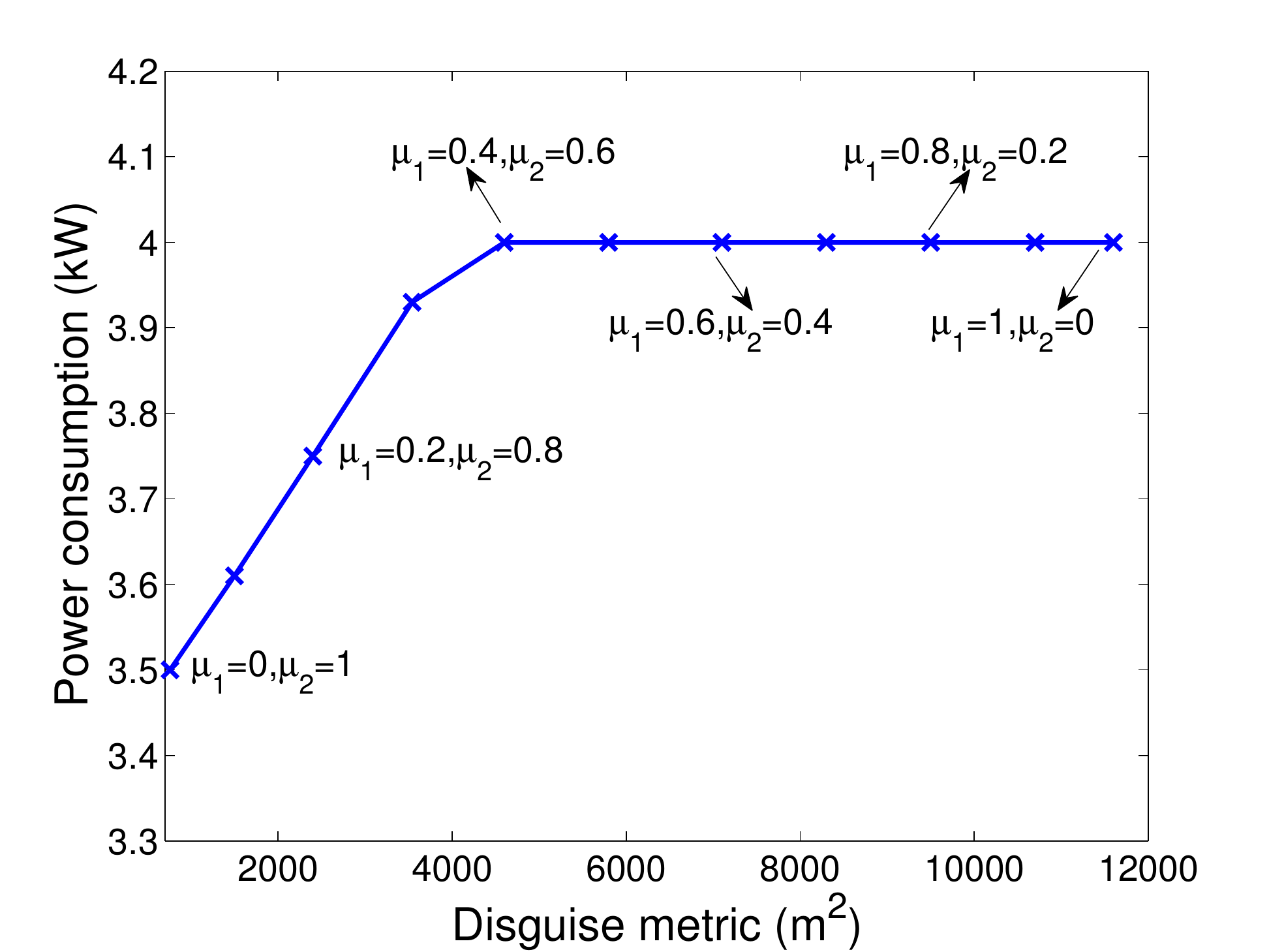}
\caption{Power consumption versus disguising metric with $\mu_1 + \mu_2 =1$.}
\label{pareto} 
\end{figure}

It is revealed in Figs. \ref{horizon6} and \ref{height6} that the MDR scheme does a good job in keeping distance
from the target and changing the heading of the UAV at the same time on the horizontal plane.
The Proposed scheme strikes the best balance between distance keeping and altitude changing,
while the baseline schemes can only deal with either distance or altitude.
Although the monitor lags behind the target under the Online scheme, its trajectory is close to and aligns with the target trajectory.
The monitor flies to exactly the same location projected on the $(x,y)$-plane with a few meters above the target from time to time under the ACO scheme.
The monitor always travels at the lowest height under the DKO scheme,
and its horizontal trajectory aligns with the target trajectory from the $10$th to the $18$th second.
Under the NDP scheme, the monitor stays close to the target horizontally and at a constant height for almost the entire scheduling period.
All the aforementioned baseline schemes have their drawbacks which put the target on the alert
and place the monitor under the risk of exposure.
Besides, altitude changing is much more frequent than desired for the ACO scheme,
which may incur the instability of the monitor and damage the generator.
In a nutshell, these simulation results fully demonstrate and corroborate the merits of our proposed schemes.

Last but not least, to demonstrate the energy saving of the proposed framework under a practical model for propulsion power,
we consider a baseline scheme (labeled as ``DST'') where the propulsion power is simply modeled as the total 2D distance traveled during the surveillance, i.e., $P_h' := \sum_t \sqrt{(x_t-x_{t-1})^2+(y_t-y_{t-1})^2}$~\cite{hailong20}.
With ${P_h'}^2$ replacing the original $\sum_t P_h^t$ in \eqref{p1},
Fig. \ref{save} depicts the corresponding monitor trajectory and power consumption.
It can be evidently seen that the proposed schemes result in a more energy-efficient trajectory,
and allows the monitor to consume less energy.
A total energy of $1424$ Joules can be saved with the proposed offline scheme.
{\blue%
It is also shown in Fig. \ref{save} that the online scheme is more energy-efficient than the DST scheme,
given its objective of energy minimization (as opposed to the travel distance minimization objective of the DST scheme).
Yet, the proposed online scheme consumes more energy than its offline counterpart,
since the offline scheme assumes the a-priori knowledge of the target's trajectory throughout the video surveillance mission.
On the other hand, the monitor keeps the target within its sight.
As the target keeps changing its speed and heading, the monitor keeps adjusting its speed and heading accordingly.
It can be seen in \eqref{speed} that the propulsion power is a function of the UAV speed.
Therefore, the time-varying propulsion power shown in Fig. \ref{save} also indicates that the UAV speed changes over time.
}

Fig. \ref{covdis} depicts the disguising performance and 3D monitor-target distance of the proposed scheme over time.
It can be observed that the disguising metric decreases when the monitor flies near the target.
The monitor keeps the maximum allowed 3D distance of $D=20$ m for effective video surveillance
from the target in most of the time.
Fig. \ref{pareto} plots the Pareto-front between the power consumption and the disguising metric
by varying the values of $\mu_1$ and $\mu_2$ ($\mu_1+\mu_2=1$).
It is seen that the power consumption does not increase anymore when the disguising metric exceeds $4600$~(m$^2$).

\section{Conclusion}\label{sec.con}
In this paper, a new framework was proposed to exploit the flexibility of a solar-powered UAV for covert video surveillance
by joint power management and 3D trajectory optimization.
The SCA and PDCAE techniques were leveraged
to convexify the optimization problem and obtain a low-complexity suboptimal solution.
The proposed PDCAE-based solution can serve as the control model to operate online
and achieve control-based refinement of the monitor's trajectory.
Extensive numerical results corroborated the merits of the proposed scheme in maintaining covertness
over baseline schemes with partial or no disguising.



\bibliographystyle{IEEEtran}
\bibliography{covert_ref}

\begin{thebibliography}{10}
\providecommand{\url}[1]{#1}
\csname url@samestyle\endcsname
\providecommand{\newblock}{\relax}
\providecommand{\bibinfo}[2]{#2}
\providecommand{\BIBentrySTDinterwordspacing}{\spaceskip=0pt\relax}
\providecommand{\BIBentryALTinterwordstretchfactor}{4}
\providecommand{\BIBentryALTinterwordspacing}{\spaceskip=\fontdimen2\font plus
\BIBentryALTinterwordstretchfactor\fontdimen3\font minus
  \fontdimen4\font\relax}
\providecommand{\BIBforeignlanguage}[2]{{%
\expandafter\ifx\csname l@#1\endcsname\relax
\typeout{** WARNING: IEEEtran.bst: No hyphenation pattern has been}%
\typeout{** loaded for the language `#1'. Using the pattern for}%
\typeout{** the default language instead.}%
\else
\language=\csname l@#1\endcsname
\fi
#2}}
\providecommand{\BIBdecl}{\relax}
\BIBdecl

\bibitem{alota19}
E.~T. {Alotaibi}, S.~S. {Alqefari}, and A.~{Koubaa}, ``{LSAR: Multi-UAV}
  collaboration for search and rescue missions,'' \emph{IEEE Access}, vol.~7,
  pp. 55\,817--55\,832, 2019.

\bibitem{zhou18}
Z.~{Zhou}, C.~{Zhang}, C.~{Xu}, F.~{Xiong}, Y.~{Zhang}, and T.~{Umer},
  ``Energy-efficient industrial internet of {UAV}s for power line inspection in
  smart grid,'' \emph{IEEE Trans. Ind. Informat.}, vol.~14, no.~6, pp.
  2705--2714, Jun. 2018.

\bibitem{chi19}
C.~Yuan, Z.~Liu, and Y.~Zhang, ``Learning-based smoke detection for unmanned
  aerial vehicles applied to forest fire surveillance,'' \emph{J. Intell.
  Robot. Syst.}, vol.~93, no.~1, pp. 337--349, Feb. 2019.

\bibitem{wang19parcel}
D.~{Wang}, P.~{Hu}, J.~{Du}, P.~{Zhou}, T.~{Deng}, and M.~{Hu}, ``Routing and
  scheduling for hybrid truck-drone collaborative parcel delivery with
  independent and truck-carried drones,'' \emph{IEEE Internet Things J.},
  vol.~6, no.~6, pp. 10\,483--10\,495, Dec. 2019.

\bibitem{zeng19}
Y.~Zeng, Q.~Wu, and R.~Zhang, ``Accessing from the sky: {A} tutorial on {UAV}
  communications for {5G} and beyond,'' \emph{Proc. IEEE}, vol. 107, no.~12,
  pp. 2327--2375, Dec. 2019.

\bibitem{kli16}
K.~Li, W.~Ni, X.~Wang, R.~Liu, S.~Kanhere, and S.~Jha, ``Energy-efficient
  cooperative relaying for unmanned aerial vehicles,'' \emph{IEEE Trans. Mobile
  Comput.}, vol.~15, no.~6, pp. 1377--1386, Jun. 2016.

\bibitem{tang19}
J.~{Tang}, G.~{Chen}, and J.~P. {Coon}, ``Secrecy performance analysis of
  wireless communications in the presence of {UAV} jammer and randomly located
  {UAV} eavesdroppers,'' \emph{IEEE Trans. Inf. Forensics Security}, vol.~14,
  no.~11, pp. 3026--3041, Nov. 2019.

\bibitem{yuan20}
X.~{Yuan}, Z.~{Feng}, W.~{Ni}, R.~P. {Liu}, J.~A. {Zhang}, and W.~{Xu},
  ``Secrecy performance of terrestrial radio links under collaborative aerial
  eavesdropping,'' \emph{IEEE Trans. Inf. Forensics Security}, vol.~15, pp.
  604--619, 2020.

\bibitem{hu20}
S.~Hu, Q.~Wu, and X.~Wang, ``Energy management and trajectory optimization for
  {UAV}-enabled legitimate monitoring systems,'' [Online]. Available:
  https://arxiv.org/abs/2004.10918, Apr. 2020.

\bibitem{yang19}
Z.~Yang, C.~Pan, K.~Wang, and M.~Shikh-Bahaei, ``Energy efficient resource
  allocation in {UAV}-enabled mobile edge computing networks,'' \emph{IEEE
  Trans. Wireless Commun.}, vol.~18, no.~9, pp. 4576--4589, Sep. 2019.

\bibitem{hailong20tii}
H.~Huang and A.~Savkin, ``An algorithm of reactive collision free 3-{D}
  deployment of networked unmanned aerial vehicles for surveillance and
  monitoring,'' \emph{IEEE Trans. Ind. Informat.}, vol.~16, no.~1, pp.
  132--140, Jan. 2020.

\bibitem{fanid20}
A.~{Alipour-Fanid}, M.~{Dabaghchian}, N.~{Wang}, P.~{Wang}, L.~{Zhao}, and
  K.~{Zeng}, ``Machine learning-based delay-aware {UAV} detection and operation
  mode identification over encrypted {Wi-Fi} traffic,'' \emph{IEEE Trans. Inf.
  Forensics Security}, vol.~15, pp. 2346--2360, 2020.

\bibitem{zhang20}
X.~{Zhang}, Y.~{Fang}, X.~{Zhang}, J.~{Jiang}, and X.~{Chen}, ``A novel
  geometric hierarchical approach for dynamic visual servoing of quadrotors,''
  \emph{IEEE Trans. Ind. Electron.}, vol.~67, no.~5, pp. 3840--3849, May 2020.

\bibitem{hailong20}
H.~Huang, A.~Savkin, and W.~Ni, ``A method for covert video surveillance of a
  car or a pedestrian by an autonomous aerial drone via trajectory planning,''
  in \emph{Proc. IEEE ICCAR}, Singapore, Apr. 2020, pp. 1--3.

\bibitem{hailong19cn}
H.~Huang and A.~Savkin, ``Reactive 3{D} deployment of a flying robotic network
  for surveillance of mobile targets,'' \emph{Comput. Netw.}, vol. 161, pp.
  172--182, Oct. 2019.

\bibitem{huang16}
Y.~Huang, H.~Wang, and P.~Yao, ``Energy-optimal path planning for solar-powered
  {UAV} with tracking moving ground target,'' \emph{Aerospace Sci. Tech.},
  vol.~53, pp. 241--251, Jun. 2016.

\bibitem{zhang18tie}
L.~{Zhang}, F.~{Deng}, J.~{Chen}, Y.~{Bi}, S.~K. {Phang}, X.~{Chen}, and B.~M.
  {Chen}, ``Vision-based target three-dimensional geolocation using unmanned
  aerial vehicles,'' \emph{IEEE Trans. Ind. Electron.}, vol.~65, no.~10, pp.
  8052--8061, Oct. 2018.

\bibitem{yao19}
P.~{Yao}, Z.~{Xie}, and P.~{Ren}, ``Optimal {UAV} route planning for coverage
  search of stationary target in river,'' \emph{IEEE Trans. Control Syst.
  Tech.}, vol.~27, no.~2, pp. 822--829, Mar. 2019.

\bibitem{shafer08}
V.~Shaferman and T.~Shima, ``Unmanned aerial vehicles cooperative tracking of
  moving ground target in urban environments,'' \emph{J. Guid. Control Dyn.},
  vol.~31, no.~5, pp. 1360--1371, 2008.

\bibitem{yu15}
H.~{Yu}, K.~{Meier}, M.~{Argyle}, and R.~W. {Beard}, ``Cooperative path
  planning for target tracking in urban environments using unmanned air and
  ground vehicles,'' \emph{IEEE/ASME Trans. Mechatronics}, vol.~20, no.~2, pp.
  541--552, Apr. 2015.

\bibitem{quintero14}
S.~A. Quintero and J.~P. Hespanha, ``Vision-based target tracking with a small
  {UAV}: Optimization-based control strategies,'' \emph{Control Engineer.
  Practice}, vol.~32, pp. 28--42, 2014.

\bibitem{brizon}
S.~Schopferer, M.~Brizon, C.~Liersch, and S.~Froese, ``Evaluating the energy
  balance of high altitude platforms at early design stages,'' in \emph{Proc.
  Int. Conf. Unmanned Aircraft Syst.}, Arlington, VA USA, Jun. 2016.

\bibitem{aglie}
G.~S. Aglietti, S.~Redi, A.~R. Tatnall, and T.~Markvart, ``Harnessing
  high-altitude solar power,'' \emph{IEEE Trans. Energy Conver.}, vol.~24,
  no.~2, pp. 442--451, Jun. 2009.

\bibitem{derrick}
Y.~Sun, D.~W.~K. Ng, D.~Xu, L.~Dai, and R.~Schober, ``Optimal 3{D}-trajectory
  design and resource allocation for solar-powered {UAV} communication
  systems,'' \emph{IEEE Trans. Commun.}, vol.~67, no.~6, pp. 4281--4298, Jun.
  2019.

\bibitem{hua}
M.~Hua, L.~Yang, Q.~Wu, and A.~L. Swindlehurst, ``3{D UAV} trajectory and
  communication design for simultaneous uplink and downlink transmission,''
  \emph{IEEE Trans. Commun.}, vol.~68, no.~9, pp. 5908--5923, Sep. 2020.

\bibitem{wmfeng}
W.~{Feng}, J.~{Tang}, Y.~{Yu}, J.~{Song}, N.~{Zhao}, G.~{Chen}, K.~K. {Wong},
  and J.~{Chambers}, ``{UAV}-enabled {SWIPT in IoT} networks for emergency
  communications,'' \emph{IEEE Wireless Commun.}, vol.~27, no.~5, pp. 140--147,
  Oct. 2020.

\bibitem{cai}
Y.~Cai, Z.~Wei, R.~Li, D.~W.~K. Ng, and J.~Yuan, ``Joint trajectory and
  resource allocation design for energy-efficient secure {UAV} communication
  systems,'' \emph{IEEE Trans. Commun.}, vol.~68, no.~7, pp. 4536--4553, Jul.
  2020.

\bibitem{ruide}
R.~Li, Z.~Wei, L.~Yang, D.~W.~K. Ng, J.~Yuan, and J.~An, ``Resource allocation
  for secure multi-{UAV} communication systems with multi-eavesdropper,''
  \emph{IEEE Trans. Commun.}, vol.~68, no.~7, pp. 4490--4506, Jul. 2020.

\bibitem{ssfang}
S.~{Fang}, G.~{Chen}, and Y.~{Li}, ``Joint optimization for secure intelligent
  reflecting surface assisted {UAV} networks,'' \emph{IEEE Commun. Lett.},
  Early access, Sep. 2020.

\bibitem{zeng16}
Y.~Zeng, R.~Zhang, and T.~J. Lim, ``Throughput maximization for {UAV}-enabled
  mobile relaying systems,'' \emph{IEEE Trans. Commun.}, vol.~64, no.~12, pp.
  4983--4996, Dec. 2016.

\bibitem{zyong}
Y.~Zeng, J.~Xu, and R.~Zhang, ``Energy minimization for wireless communication
  with rotary-wing {UAV},'' \emph{IEEE Trans. Wireless Commun.}, vol.~18,
  no.~4, pp. 2329--2345, Apr. 2019.

\bibitem{seddon11}
J.~M. Seddon and S.~Newman, \emph{Basic Helicopter Aerodynamics}, 3rd~ed.\hskip
  1em plus 0.5em minus 0.4em\relax Hoboken, NJ, USA: Wiley, 2011.

\bibitem{liucamera}
H.~Liu, S.~Chen, and N.~Kubota, ``Intelligent video systems and analytics: A
  survey,'' \emph{IEEE Trans. Ind. Informat.}, vol.~9, no.~3, pp. 1222--1233,
  Aug. 2013.

\bibitem{schramm}
S.~Schramm, J.~Rangel, D.~A. Salazar, R.~Schmoll, and A.~Kroll, ``Target
  analysis for the multispectral geometric calibration of cameras in visual and
  infrared spectral range,'' \emph{IEEE Sensors J.}, Aug. 2020.

\bibitem{Boyd}
S.~Boyd and L.~Vandenberghe, \emph{Convex Optimization}.\hskip 1em plus 0.5em
  minus 0.4em\relax Cambridge University Press, 2004.

\bibitem{sca18}
Y.~{Yang}, M.~{Pesavento}, S.~{Chatzinotas}, and B.~{Ottersten}, ``Successive
  convex approximation algorithms for sparse signal estimation with nonconvex
  regularizations,'' \emph{IEEE J. Sel. Topics Signal Process.}, vol.~12,
  no.~6, pp. 1286--1302, Dec. 2018.

\bibitem{wen18}
B.~Wen, X.~Chen, and T.~K. Pong, ``A proximal difference-of-convex algorithm
  with extrapolation,'' \emph{Comput. Optim. Appl.}, vol.~69, no.~2, pp.
  297--324, Mar. 2018.

\bibitem{gotoh18}
J.-Y. Gotoh, A.~Takeda, and K.~Tono, ``D{C} formulations and algorithms for
  sparse optimization problems,'' \emph{Math. Program.}, vol. 169, no.~1, pp.
  141--176, May 2018.

\bibitem{nest07}
Y.~Nesterov, ``Dual extrapolation and its applications to solving variational
  inequalities and related problems,'' \emph{Math. Program.}, vol. 109, no.~2,
  pp. 319--344, Mar. 2007.

\bibitem{nest13}
------, ``Gradient methods for minimizing composite functions,'' \emph{Math.
  Program.}, vol. 140, no.~1, pp. 125--161, Aug. 2013.

\bibitem{mpc13}
G.~C. {Calafiore} and L.~{Fagiano}, ``Robust model predictive control via
  scenario optimization,'' \emph{IEEE Trans. Autom. Control}, vol.~58, no.~1,
  pp. 219--224, Jan. 2013.

\bibitem{kokh04}
A.~Kokhanovsky, ``Optical properties of terrestrial clouds,''
  \emph{Earth-Science Reviews}, vol.~64, no.~3, pp. 189--241, Feb. 2004.

\end{thebibliography}
\end{document}